\documentclass[12pt]{article}
\usepackage{latexsym}
\usepackage{graphicx}
 \renewcommand{\baselinestretch}{1.5}
\newcommand{\Sigmavec}{\mbox{\boldmath $\Sigma$}}

\newtheorem{rema}{\hskip\parindent\sc Remark}
\newtheorem{theo}{\hskip\parindent\sc Theorem}
\newtheorem{coll}{\hskip\parindent\sc Corollary}

 \makeatletter
 \@addtoreset{equation}{section}
 \makeatother

\begin{document}
\catcode`@=11
\newskip\plaincentering \plaincentering=0pt plus 1000pt minus 1000pt
\def\plainLet@{\relax\iffalse{\fi\let\\=\cr\iffalse}\fi}
\def\plainvspace@{\def\vspace##1{\noalign{\vskip##1}}}
\newif\iftagsleft@
\tagsleft@true
\def\TagsOnRight{\global\tagsleft@false}
\TagsOnRight
\def\tag#1$${\iftagsleft@\leqno\else\eqno\fi
 \hbox{\def\pagebreak{\global\postdisplaypenalty-\@M}%
 \nopagebreak{\global\postdisplaypenalty\@M}\rm(#1\unskip)}%
  $$\postdisplaypenalty\z@\ignorespaces}
\interdisplaylinepenalty=\@M
\def\plainallowdisplaybreak@{\def\allowdisplaybreak{\noalign{\allowbreak}}}
\def\plaindisplaybreak@{\def\displaybreak{\noalign{\break}}}
\def\align#1\endalign{\def\tag{&}\plainvspace@\plainallowdisplaybreak@\plaindisplaybreak@
  \iftagsleft@\plainlalign@#1\endalign\else
   \plainralign@#1\endalign\fi}
\def\plainralign@#1\endalign{\displ@y\plainLet@\tabskip\plaincentering\halign to\displaywidth
     {\hfil$\displaystyle{##}$\tabskip=\z@&$\displaystyle{{}##}$\hfil
       \tabskip=\plaincentering&\llap{\hbox{(\rm##\unskip)}}\tabskip\z@\crcr
             #1\crcr}}
\def\plainlalign@
 #1\endalign{\displ@y\plainLet@\tabskip\plaincentering\halign to \displaywidth
   {\hfil$\displaystyle{##}$\tabskip=\z@&$\displaystyle{{}##}$\hfil
   \tabskip=\plaincentering&\kern-\displaywidth
        \rlap{\hbox{(\rm##\unskip)}}\tabskip=\displaywidth\crcr
               #1\crcr}}
\catcode`@=12

\def\sect{\section{\leftline{\large\bf}}}
\thispagestyle{empty}
\vspace*{-13mm}

\begin{center}
{\large\bf ESTIMATION FOR A PARTIAL-LINEAR\\ SINGLE-INDEX MODEL}
\\[0.5cm]
{\large Jane-Ling Wang${^1}$, Liugen Xue${^2}$, Lixing Zhu$^{3}$,
and Yun Sam Chong${^4}$
} \\[0.4cm]
{\it $^1$University of California at Davis}\\
{\it $^2$Beijing University of Technology, Beijing, China }\\
{\it $^3$Hong Kong Baptist University, Hong Kong, China} \\
{\it $^4$Wecker Associate}
\end{center}
\vspace{0.2cm}

\begin{center}
\renewcommand{\baselinestretch}{1.5}
\parbox{0.8\hsize}
{~~~~\small In this paper, we study the estimation for a
partial-linear single-index model. A two-stage estimation
procedure is proposed to estimate the link function for the single
index and the parameters in the single index, as well as the
parameters in the linear component of the model. Asymptotic
normality is established for both parametric components. For the
index, a constrained  estimating equation leads to an
asymptotically more efficient estimator than  existing estimators
in the sense that it is of a smaller limiting variance. The
estimator of the nonparametric link function achieves optimal
convergence rates; and the structural error variance is obtained.
In addition, the results facilitate the construction of confidence
regions and hypothesis testing for the unknown parameters.  A
simulation study is performed and an application to a real dataset
is illustrated. The extension to multiple indices is briefly
sketched.}
\end{center}
\vspace{0.2cm}

%
%

{
\renewcommand{\baselinestretch}{1}
 \footnotetext{ Lixing Zhu is the corresponding author. Email:
lzhu@hkbu.edu.hk.  Jane-Ling Wang's research was partially supported
by NSF grant DMS-0406430. Liugen Xue's research was supported by the
National Natural Science Foundation of China (10571008, 10871013),
the Natural Science Foundation of Beijing (1072004) and Ph. D.
Program Foundation of Ministry of Education of China (20070005003).
Lixing Zhu's research was supported by a grant of The Research Grant
Council of Hong Kong, Hong Kong, China (HKBU7060/04P and HKBU
2030/07P). The first three authors have equal contribution to this
research. The authors thank the Editor, the Associate Editor, and
the two referees for their insightful comments and suggestions which
have led to substantial improvements in the presentation of the
manuscript.}

 \footnotetext{ {\it AMS 2000 subject classifications}. Primary 62G05; secondary
 62G20.}

 \footnotetext{ {\it Key words and phrases}. Dimension reduction, local linear
smoothing, bandwidth, two-stage estimation, kernel smoother.}
 }

\newpage

\section{Introduction }

\hskip\parindent
 Partial linear models have  attracted lots of
attention due to their flexibility to combine traditional linear
models with nonparametric regression models. See, e.g. Heckman
(1986), Rice (1986),  Chen (1988), Bhattacharya and Zhao (1997),
Xia and H\"ardle (2006), and the recent comprehensive  books by
H\"ardle, Gao, and Liang (2000) and Ruppert, Wand and Carroll
(2003) for additional references. However, the nonparametric
components are subject to the curse of dimensionality and can only
accommodate low dimensional covariates $X$.  To remedy this, a
dimension reduction model which assumes that the influence of the
covariate $X$ can be collapsed to a single index, $X^{\rm T}
\beta$, through a nonparametric link function $g$ is a viable
option and termed the partial-linear single-index model.
Specifically, it takes the form:
\begin{equation}\label{(1)}
Y=Z^{\rm T}\theta_0 + g(X^{\rm T}\beta_0)+e,
\end{equation}
where $(X, Z)\in R^p\times R^q$ are covariates of the response
variable $Y$, $g$ is an unknown link function for the single index,
and $e$ is the error term with $E(e)=0$ and $0<{\rm Var}(e)=\sigma^2
< \infty$. For the sake of identifiability, it is often assumed that
$\|\beta_0 \|=1$ and the $r$th component of $\beta_0$ is positive,
where $\|\cdot\|$ denotes the Euclidean metric.

This model is quite general, it includes the aforementioned
partial-linear model when the dimension of $X$ is one and also the
popular single-index model in the absence of the linear covariate
$Z$.  There is an extensive literature for the single-index model
with three main approaches:  projection pursuit regression (PPR)
[Friedman and Stuetzle (1981), Hall (1989), H\"ardle, Hall and
Ichimura (1993)];  the average derivative approach [Stoker (1986),
Doksum and Samarov (1995), and Hristache, Juditsky and Spokoiny
(2001)]; and  sliced inverse regression (SIR) and related methods
[Li (1991),  Cook and Li (2002), Xia, Tong, Li and Zhu (2002), and
Yin and Cook (2002)]. All these approaches rely on the assumption
that the predictors in $X$ are continuous variables, while model
(\ref{(1)}) compensates for this by allowing discrete or other
continuous variables to be linearly associated with the response
variable. To our knowledge, Carroll, Fan, Gijbels and Wand (1997)
were the first to explore model (\ref{(1)}) and they actually
considered a generalized version, where a known link function is
employed in the regression function while model (\ref{(1)}) assumes
an identity link function. However, their approaches may become
computationally unstable as observed by Yu and Ruppert (2002) and
confirmed by our simulations in Section 3. The theory of Carroll,
Fan, Gijbels and Wand (1997) also relies on the strong assumption
that their estimator for $\theta_0$ is already
$\sqrt{n}$-consistent. Yu and Ruppert (2002) alleviated both
difficulties by employing a link function $g$ which falls in a
finite-dimensional spline space, yielding essentially a flexible
parametric model. Xia and H\"ardle (2006) used a method that is
based on a local polynomial smoother and a modified version of least
squares in H\"ardle, Hall and Ichimura (1993).

In this paper, we propose a new estimation procedure.  Our approach
requires no iteration and works well under the mild condition that a
few indices based on $X$ suffice to explain $Z$. Namely,
\begin{equation}
Z = \phi(X^{\rm T}\beta_Z) + \eta, \label{(2)}
\end{equation}
where $\phi(\cdot)$ is an unknown function from $R^d$ to $R^q$,
$\beta_Z$ is a $p\times d$ matrix with orthonormal columns, $\eta$
has mean
zero and is independent of $X$. 
The dimension $d$ is often much smaller than the dimension $p$ of
$X$. Such an assumption is not stringent and common in most
dimension reduction approaches in the literature. A theoretical
justification is provided in Li, Wen and Zhu (2008). Model
(\ref{(2)}) implies that a few indices of $X$ suffice to summarize
all the information carried in $X$ to predict $Z$, which is often
the case in reality, such as for the Boston Housing data in section
4, where a single index was selected for model (1.2) and $Z$ is a
discrete variable. In this data, first analyzed in Harrison and
Rubinfeld (1978), the response variable is the median value of
houses in $506$ census tracts in the Boston area. The covariates
include: average number of rooms, the proportion of houses built
before 1940, eight variables describing the neighborhood, two
variables describing the accessibility to highways and employment
centers, and two variables describing air pollution. A key covariate
of interest is a binary variable that specifies whether a house
borders the river or not. Our analysis presented in Section 4 based
on the dimension reduction assumptions of (\ref{(1)}) and with $Z$
equal to this binary variable in (\ref{(2)}) demonstrates the
advantages of our model assumption, only one index ($d=1$) was
needed in model (\ref{(2)}) for this
 data.

To avoid the computational complications that we experienced with
the procedure in Carroll et al. (1997), who aim at estimating
$\beta_0$ and $\theta_0$ simultaneously, we choose to estimate
$\beta_0$ and $\theta_0$ sequentially. The idea is simple:
$\theta_0$ can be estimated optimally through approaches developed
for partial linear models once we have a $\sqrt{n}$ estimate of
$\beta_0$ and plug it in (1.1). However, $\beta_0$ and $\theta_0$
may be correlated, leading to difficulties in identifying
$\beta_0$. This is where model (1.2) comes in handy, as it allows
us to remove the part of $Z$ that is related to $X$ so that the
residual $\eta$ in (1.2) is independent of $X$.
Again, we need to impose the
identifiability condition that $\beta_Z$ has norm one and a positive
first component. The procedure is as follows: First estimate
$\beta_Z$ via any dimension reduction approach, such as SIR or PPR
for $q=1$, and the projective resampling method in Li, Wen and Zhu
(2008) for $q>1$. Once $\beta_Z$ has been estimated we proceed to
estimate $\phi$ via a $d$-dimensional smoother  and then obtain the
residual for $\eta$.
 Since $\eta=Z-\phi(X^{\rm T}\beta_Z)$, plugging this into
(\ref{(1)}) we get
$$
Y=\eta^{\rm T}\theta_0+ h(X^{\rm T}\beta_0,X^{\rm T}\beta_Z)+e,
$$
where $h$ is an unknown function, but now $\eta$ and $X$ are
independent of each other.  It is thus possible to employ a least
squares approach to estimate $\theta_0$ and the resulting estimate
will be $\sqrt{n}$-consistent.  
We then employ a dimension reduction procedure to $Y-Z^{\rm T}
\hat{\theta}_0$ and $X$ to obtain an estimate for $\beta_0$ and
$g$. This concludes the first stage , where the resulting
estimates for $\theta_0$ and $\beta_0$ are already $\sqrt{n}$
consistent but will serve the role as initial estimates for the
next stage, where we update all the estimates but use a more
sophisticated approach. Specifically for $\theta_0$ we apply the
profile method, also called partial regression in Speckman (1988),
to estimate $\theta_0$. Theoretical results in Section 2.2
indicate that the two-stage procedure is fully efficient, so there
is no need for iteration. More importantly, to estimate the index
$\beta_0$, we use an estimating equation to obtain asymptotic
normality,  which takes the constraint $\|\beta_0\|=1$ into
account. The estimator based on this new estimating equation
performs better in several ways,  summarized as follows.

\begin{enumerate}
\item Our estimation procedure directly targets the model
parameters
 $\theta_0$, $\beta_0$, $\beta_Z$, $\phi(\cdot)$ and $g(\cdot)$ and no iteration is needed.

\item We obtain the asymptotic normality of the estimator of
$\beta_0$ and the optimal convergence rate of the estimator of
$g(\cdot)$, as well as the asymptotic normality of the estimator of
$\theta_0$. The most attractive feature  of this new method is that
the estimator of $\beta_0$ has smaller limiting variance when
compared to  three existing approaches in : H\"ardle et al. (1993)
when the model is reduced to the single-index model, Carroll et
al.(1997) if their link function is the identity function, and Xia
and H\"ardle (2006) when their model is homoscedastic. This is the
first result providing such a small limiting variance in this area.



\item We also provide the asymptotic normality of the estimator of
$\sigma^2$. It allows us to consider the construction of confidence
regions and hypothesis testing for $\theta_0$ and $\beta_0$.

\end{enumerate}

The rest of the paper is organized as follows. In Section 2, we
elaborate on the new methodology and then present the asymptotic
properties for the estimators. Section 3 reports the results of a
simulation study and Section 4  an application to a real data
example for illustration.
 Section 5 gives the proofs of the main theorems. Some
lemmas and their proofs are relegated to the Appendix.

\section{Methodology and Main Results}

\subsection{Estimating Procedures }

 \hskip\parindent
The observations are $\big\{(X_i,Y_i,Z_i);1\leq i\leq n\big\}$, a
sequence of independent and identically distributed (i.i.d.) samples
from (\ref{(1)}), i.e.
$$
Y_i=Z_i^{\rm T}\theta_0 + g(X_i^{\rm T}\beta_0)+e_i, \ \
i=1,\ldots,n,
$$
where $e_1,\cdots,e_n$ are i.i.d. random errors with $E(e_i)=0$
and ${\rm Var}(e_i)=\sigma^2>0$, $\big\{\varepsilon_i;1\leq i\leq
n\big\}$ are independent of $\big\{(X_i,Z_i);1\leq i\leq n\big\}$,
$X_i=(X_{i1},\ldots,X_{ip})^{\rm T}$,
$Z_i=(Z_{i1},\ldots,Z_{iq})^{\rm T}$, $\beta_0\in R^p$ and
$\theta_0\in R^q$. For simplicity of presentation, we initially
assume that $Z$ can be recovered from a single-index of $X$. That
is,  $d=1$ in (\ref{(2)}). The general case will be explored at
the end of this section in Remarks 2. Below we first outline  the
steps for each stage and then elaborate on each of these steps.
\\

 \textbf{Algorithm for Stage One:}
\begin{enumerate}
\item Apply a dimension reduction method for the regression of
$Z_i$ versus $X_i$ to find an estimator $\hat \beta_Z$ of
$\beta_Z$;

\item  Smooth the $Z_i$ over $\hat X_i^{\rm T}\beta_Z$ to get an
estimator $\hat \phi(\cdot)$ of $\phi(\cdot)$, then compute the
residuals $\hat \eta_i=Z_i-\hat \phi(X_i^{\rm T}\hat \beta_Z)$;

\item Perform a linear regression of $Y_i$ versus $\hat \eta_i$'s
to find an initial estimator $\hat \theta_0$ of $\theta_0$;

\item Apply a dimension reduction method to the regression of
$Y_i-Z_i^{\rm T}\hat \theta_0$ versus $X_i$  to find an initial
estimator $\hat \beta_0$ of $\beta_0$

\item Smooth the $Y_i-Z_i^{\rm T}\hat\theta_0$ versus the
$X_i^{\rm T} \hat \beta_0$  to obtain an estimator for $g$ and for
its derivative $g'$. \\

\noindent \mbox{\textbf{Algorithm for Stage Two:}}

\item  Use the initial estimate $\hat \beta_0$ from Step 4 to
update the estimate of $\theta_0$ through a profile approach for
the partial linear model by minimizing (\ref{PR}).

\item Use the updated  estimate $\hat \theta$ of $\theta_0$ from
Step 6 to form the new residual $Y-Z^{\rm T}\hat \theta$, then
update the estimate of $\beta_0$ by solving the estimating
equation (\ref{beta}).

\item Use the updated estimates of $\theta_0$ and $\beta_0$ in
Steps 6 and 7 to update the estimate of $g$, following  the
procedure as described in Step 5.
\end{enumerate}

This completes the algorithm and, as we show in Section 2.2,  the
resulting estimators are already theoretically efficient. However,
the practical performance can be improved by iterating Steps 6 and
7 one or more times. Our experience, through simulation studies
not reported in this paper, reveals limited benefits when
iterating more than once.

Next, we elaborate on each of the steps in the above algorithms for
the simple case of a single index ($d=1$). For the dimension
reduction method in Step 4, one can use any of several existing
methods, such as SIR or one of its variants, PPR, or the minimum
average variance estimator (MAVE) of Xia, Tong, Li and Zhu (2002).
These methods are for univariate responses and hence can also be
applied in Step 1 when $q=1$. However, when $q>1$, a different
method is needed in Step 1 for the case of a  multivariate response,
and we recommend the dimension reduction method in Li, Wen and Zhu
(2008). This and other results in the literature already demonstrate
the $\sqrt{n}$-consistency of these dimension reduction methods.

For the smoothing involved in Step 5, one can choose any
one-dimensional smoother.  We employ the local polynomial smoother
(Fan and Gijbels, 1996) to obtain estimators of the link function
 $g$ and its derivative $g'$, which will be used in the second stage
of the estimation procedure. Specifically, for a kernel function
$K(\cdot)$ on $R^1$ and a bandwidth sequence $b=b_n$, define
$K_b(\cdot)=b^{-1}K(\cdot/b)$. For a fixed $\beta$ and $\theta$, the
 local linear smoother aims at minimizing the weighted sum of squares
$$
\sum_{i=1}^n\big[Y_i-Z_i^{\rm T}\theta-d_0-d_1(X_i^{\rm
T}\beta-t)\big]^2K_b(X_i^{\rm T}\beta-t)
$$
with respect to the parameters $d_\nu$, $\nu=0, 1$. Let $h=h_n$ and
$h_1=h_{1n}$ denote the bandwidths for estimating $g(\cdot)$ and
$g'(\cdot)$, respectively. A simple calculation shows that the local
linear smoother with these specifications can be represented  as
\begin{equation}
 \hat{g}(t;\beta,\theta) =\sum_{i=1}^nW_{ni}(t,\beta)(Y_i-Z_i^{\rm T}
 \theta),
\label{ghat} \end{equation} and \begin{equation}
 \hat{g}'(t;\beta,\theta) =\sum_{i=1}^n\widetilde{W}_{ni}(t,\beta)
 (Y_i-Z_i^{\rm T}\theta),
\label{g'hat}\end{equation} where
\begin{equation}
W_{ni}\big(t;\beta\big) =\frac{K_h\big(X_i^{\rm
T}\beta-t\big)\big[S_{n,2}\big(t;\beta,h\big)-(X_i^{\rm
T}\beta-t)S_{n,1}\big(t;\beta,h\big)\big]}{S_{n,0}\big(t;\beta,h\big)
S_{n,2}\big(t;\beta,h\big)-S_{n,1}^2\big(t;\beta,h\big)\big]},
 \label{(3)}
\end{equation}
\begin{equation}
\widetilde{W}_{ni}\big(t;\beta\big) =\frac{K_{h_1}\big(X_i^{\rm
T}\beta-t\big)\big[(X_i^{\rm
T}\beta-t)S_{n,0}\big(t;\beta,h_1\big)-S_{n,1}\big(t;\beta,h_1\big)
\big]}{S_{n,0}\big(t;\beta,h_1\big)S_{n,2}\big(t;\beta,h_1\big)-
S_{n,1}^2\big(t;\beta,h_1\big)\big]},
 \label{(4)}
\end{equation} and
$$
S_{n,l}\big(t;\beta,h\big)
 =\frac{1}{n}\sum_{i=1}^n\big(X_i^{\rm T}\beta-t\big)^lK_h\big
 (X_i^{\rm T}\beta-t\big),\ \ l=0, 1, 2.
$$

The above estimators are for generic fixed values of $\beta$ and
$\theta$. To obtain  the estimates needed in Step 5, one  replaces
them with the initial values $\hat{\beta}_0$ obtained in Step 1
and $\hat{\theta}_0$ obtained in Step 3, respectively. We will
show in Theorem 2 that this results in standard convergence rates
for the estimate of $g$.

Likewise, a local linear smoother can be employed in Step 2 for
estimating the unknown function $\phi$ in model (\ref{(2)}).  The
resulting estimator is defined as
$$
\hat{\phi}\big(t;\hat{\beta}_Z\big)=
\sum_{i=1}^nW_{ni}\big(t;\hat{\beta}_Z\big)Z_i.
$$


Several possibilities are available for the  estimator of $\theta_0$
in Step 6, such as the profile approach (termed ``partial
regression'' in Speckman, 1988) or the partial spline approach
(Heckman, 1986). Here the the partial spline approach is not
suitable for correlated $X$ and $Z$, so we adopt  a profile approach
and a local linear smoother. In short, this amounts to minimizing,
over all $\theta$, the sum of squared errors,
\begin{equation}  \label{PR}
 \sum_{i=1}^n\big[Y_i-Z_i^{\rm T}\theta-\hat{g}(X_i^{\rm
T}\hat{\beta}_0;\hat{\beta}_0, \theta)\big]^2,
\end{equation}
where $\hat{g}$ is the estimator in (\ref{ghat}) of $g$, obtained
by smoothing $Y_i-Z_i^{\rm T}\theta$ versus $X_i^{\rm
T}\hat{\beta}_0$, and $\hat{\beta}_0$ is an initial estimator of
$\beta_0$, which could be the initial estimator $\hat{\beta}_0$ in
Step 4 or the refined estimator from Step 7 when an iterated
estimator for $\theta_0$ is desirable. Because this smoother is
expressed as a function of $\theta$, the estimate derived from
(\ref{PR}) is a profile estimate. More details about the
derivation and advantages of the profile approach can be found in
Speckman (1988). Specifically, let $\hat{\beta}_0$ be the current
estimator,
  $\tilde{\bf Y}=(\tilde{Y}_1,\ldots,\tilde{Y}_n)^{\rm T}$,
  $\tilde{\bf Z}=(\tilde{Z}_1,\ldots,\tilde{Z}_n)^{\rm T}$,
  where
$$\begin{array}{ll}
 \tilde{Y}_i=Y_i-\hat{g}_1\big(X_i^{\rm T}\hat{\beta}_0;\hat{\beta}_0\big),
 & \tilde{Z}_i=Z_i-\hat{g}_2\big(X_i^{\rm T}\hat{\beta}_0;\hat{\beta}_0\big),
 \\
 \hat{g}_{1}\big(t;\hat{\beta}_0\big)= \sum_{i=1}^nW_{ni}\big(t;\hat{\beta}_0\big)Y_i,\ \ \ \
 & \hat{g}_2\big(t;\hat{\beta}_0\big)=
 \sum_{i=1}^nW_{ni}\big(t;\hat{\beta}_0\big)Z_i,
\end{array} $$
with $\hat{g}_1$ and $\hat{g}_2$ the respective estimators of
$g_1(t)=E\big(Y|X^{\rm T}\beta_0=t\big)$ and $g_2(t)=E\big(Z|X^{\rm
T}\beta_0=t\big)$. The resulting partial regression estimator is
thus
\begin{equation}
\hat{\theta}=(\tilde{\bf Z}^{\rm T}\tilde{\bf Z})^{-1}\tilde{\bf
Z}^{\rm T}\tilde{\bf Y}.
 \label{(5)}
\end{equation}
For the estimator of $\beta_0$ in Step 7, we propose a novel method
that takes advantage of the constraint $\|\beta_0\|=1$ and hence  is
more efficient than existing approaches, including the PPR approach
in  H\"ardle et al (1993),  the MAVE method in Xia et al. (2002),
and the least squares approaches of Carroll et al (1997) and Xia and
H\"ardle (2006) for the single-index partial linear model in
(\ref{(1)}). It is worth mentioning that Xia and H\"ardle (2006)
allow possible heteroscadestic structure in (\ref{(1)}),  and least
squares approaches have been standard dimension
 methods
 and lead to the same asymptotic variances for estimators of $\beta_0$.  For
instance, in the homoscadestic case, the estimator in Xia and
H\"ardle (2006) has an asymptotic variance that is  identical to
that of H\"ardle et al (1993). Our approach, based on an estimating
equation under the constraint $\|\beta_0\|=1$, is computationally
stable and asymptotically more efficient, i.e., its asymptotic
variance is smaller. The efficiency gain can be attributed to a
re-parametrization, making use of the constraint $\|\beta_0\|=1$ by
transferring restricted least squares to un-restricted least
squares, which makes  it possible to search for the solution of the
estimating equation over a restricted region in the Euclidean space
$R^{p-1}$.

 Without loss of generality, we may
assume that the true parameter $\beta_0$ has a positive component
(otherwise, consider $-\beta_0$), say $\beta_{0r}>0$ for
$\beta_0=(\beta_{01},\ldots,\beta_{0p})^{\rm T}$ and $1\leq r\leq
p$. For $\beta=(\beta_1,\ldots,\beta_p)^{\rm T}$, let
$\beta^{(r)}=(\beta_1,\ldots,\beta_{r-1},\beta_{r+1},\ldots,\beta_p)^{\rm
T}$ be a $p-1$ dimensional parameter vector after removing the $r$th
component $\beta_r$ in $\beta$. Then we may write
\begin{equation}
 \beta=\beta(\beta^{(r)}) = (\beta_1,\ldots,\beta_{r-1}, (1-\|\beta^{(r)}\|^2)^{1/2},\beta_{r+1},\ldots,\beta_{p})^{\rm T}.
 \label{(6)}
\end{equation}
The true parameter $\beta_0^{(r)}$ must satisfy the constraint
$\|\beta_0^{(r)}\|< 1$, and $\beta$ is infinitely differentiable in
a neighborhood of $\beta_0^{(r)}$. This ``remove-one-component"
method for $\beta$ has also been  applied in Yu and Ruppert (2002).


To obtain the estimator, consider  a  {\rm Jacobian} matrix of
$\beta$ with respect to  $\beta^{(r)}$,
\begin{equation} {\bf
J}_{\beta^{(r)}}=\frac{\partial\beta}{\partial{\beta^{(r)}}}=(\gamma_1,\ldots,
\gamma_p)^{\rm T},
 \label{(7)}
\end{equation}
where $\gamma_s$ $(1\leq s\leq p, s\neq r)$ is a $p-1$ dimensional
unit vector with $s$th component 1, and
$\gamma_r=-(1-\|\beta^{(r)}\|^2)^{-1/2}\beta^{(r)}$. To motivate the
estimating equation, we start with the least squares  criterion:
\begin{equation}
D(\beta):=\sum_{i=1}^n \big[Y_i-Z_i^{\rm
T}\hat{\theta}-\hat{g}(X_i^{\rm
T}\beta;\beta,\hat{\theta})\big]^2.
 \label{(8)}
\end{equation}
From (\ref{(6)}) and (\ref{(8)}) we find
$D(\beta)=D(\beta(\beta^{(r)}))=\tilde{D}(\beta^{(r)})$. Therefore,
we may obtain an estimator of $\beta^{(r)}_0$, say
$\hat{\beta}^{(r)}$, by minimizing $\tilde{D}(\beta^{(r)})$, and
then obtain an estimator of $\beta_0$, $\hat{\beta}$, via a
transformation. This means that we transform  a restricted least
squares problem to an unrestricted least squares problem  by solving
the estimation equation:
\begin{equation}
\sum_{i=1}^n \big[Y_i-Z_i^{\rm T}\hat{\theta}-\hat{g}(X_i^{\rm
T}\beta;\beta,\hat{\theta})\big]\hat{g}'(X_i^{\rm
T}\beta;\beta,\hat{\theta}){\bf J}_{\beta^{(r)}}^TX_i=0.
 \label{beta}
\end{equation}
We define the resulting estimator $\hat \beta$ of $\beta_0$ as the
final target estimator. Theorem 3 implies that  our estimator for
$\beta_0$ has a smaller limiting variance than the estimators in Xia
and H\"{a}rdle (2006) and Carroll et al. (1997).



With $\hat{\theta}$ and $\hat{\beta}$, the final estimator
$\hat{g}^*$ of $g$ in Step 8 can be defined by
$$
\hat{g}^*(t):=\hat{g}\big(t;\hat{\beta},\hat{\theta}\big)
 =\sum_{i=1}^nW_{ni}\big(t;\hat{\beta}\big)(Y_i-Z_i^{\rm T}\hat{\theta}),
$$
and the estimator $\hat{\sigma}^2$ of $\sigma^2$ by $
\hat{\sigma}^2=\frac{1}{n}\sum_{i=1}^n\big[Y_i-Z_i^{\rm
T}\hat{\theta}-\hat{g}^*(X_i^{\rm T}\hat{\beta})\big]^2. $
Asymptotic results for the final parameter estimates of $\theta$
and $\beta$ are established in Theorem 1 and Theorem 2, and
results for the link estimate of $g$ follow from Theorem 4.

\begin{rema}
\label{rema1}\rm\ We consider a homoscedastic model of (1.1) with
$d=1$ in model (\ref{(2)}). While the estimation procedure can be
extended easily to heteroscedastic errors, an additional dimension
reduction assumption on the variance function of of $\eta$, given
$X$, is needed to avoid the curse of high dimensional smoother
needed in Step 2 to estimate $\phi$. This assumption requires that
this variance function is also a function of a few indices based
on $X$. Moreover, the extension of asymptotic theory is not
straightforward. For instance, the asymptotic efficiency of the
estimator $\beta_0$ is technically challenging in the
heteroscedastic case and its study is beyond the scope of this
paper.
\end{rema}

\begin{rema}
\label{rema2}\rm\ So far, we have assumed that $d=1$. This
assumption can be extended without difficulty to the general case
where $d$ might be greater than 1. In this case, a multivariate
smoother will be employed for estimating $\phi(\cdot)$.
 The asymptotic results for the parameter estimates of $\beta$ and $\theta$ remain
unchanged, except that the rate of convergence for the link estimate
of  $\phi(\cdot)$ changes with the dimension of $d$.
\end{rema}

\begin{rema}
\label{rema3}\rm\ Other dimension reduction approaches, such as MAVE
(Xia et al., 2002) and other variants of SIR, such as SIR2 (Li,
1991) and SAVE (Cook and Wiseberg, 1991), could be employed in Steps
1 and 4 for the case of $q=d=1$ in (\ref{(2)}), especially when SIR
fails for the case of symmetric design of $X$. While MAVE is perhaps
the most efficient method of all, the benefits over SIR are limited,
as all estimates are updated in Stage 2, and it is in this step
where the major efficiency gains occur. In addition, MAVE is
computationally more intensive than SIR and encounters difficulties
in estimating $\beta_Z$, unless the covariate $Z$ is one-dimensional
and the dimension $d$ of $\beta_Z$ is also small. In fact, the
$\sqrt{n}$-consistency may not hold when $d>3$ in (1.2) as shown in
Xia, Tong, Li and Zhu (2002).

Also, SIR2/SAVE was shown in Li and Zhu (2007) to be not
$\sqrt{n}$-consistent, unless a bias correction is performed. In
contrast, either SIR or pHd (Li, 1992) can be employed to identify
the directions when $d>1$ and $q=1$, and both lead to
$\sqrt{n}$-consistency.
\end{rema}

\begin{rema}
\label{rema4}\rm\ When the dimension $q$ of $Z$ is greater than 1,
a multivariate extension  of SIR (Li et al., 2003) can be employed
 conceptually in Step 1 of the algorithm. However, the number of observations per
slice may become sparse, so we recommend an alternative multivariate
approach as in Li, Wen and Zhu (2008) or Zhu, Zhu, Ferr\'e and Wang
(2008) in Step 1.
\end{rema}

\begin{rema}
\label{rema5}\rm\ The single-index assumption in (\ref{(1)}) can
be easily extended to multiple indices through SIR or its
variants, but the estimation of the multivariate link function $g$
would encounter the curse of high dimensionality. Since no more
than three indices will be needed in many  applications,  the
approach in this paper can indeed  be extended in practice to
multiple indices.
\end{rema}

\subsection{Main results }



 \hskip\parindent
In this section, the $\sqrt{n}$ asymptotics for initial estimates of
$\beta_0$ and $\theta_0$ in Stage 1 are taken for granted as they
follow from existing results, so we do not formally list  the needed
assumptions for this to hold  but have provided sources after
Theorem 1 below. However, the asymptotics for the initial estimate
of $g$ and each of the parametric and nonaprametric estimates in
Stage 2 are fully developed in Section 2.2 with detailed assumptions
listed for each estimator.

In order to study the asymptotic behavior of the estimators, we
list the following  conditions:

{\parindent=0pt
\def\toto#1#2{\rightline{\hbox to 0.7cm{#1\hss}
\parbox[t]{15.7cm}{#2}}}

\toto{C1.}{(i)\ The distribution of $X$ has a compact support set $A$. \\
 (ii)\ The density function of $X^{\rm T}\beta$ is positive and satisfies a Lipschitz condition of order 1
 for $\beta$ in a neighborhood of $\beta_0$. Further, $X^{\rm T}\beta_0$ has a
positive and bounded density function $f(t)$ on ${\cal T}$, where
${\cal T}=\{t=x^{\rm T}\beta_0: x\in A\}$. }
 \vskip 10pt

\toto{C2.}{(i)\ The functions $g$ and $g_{2i}$ have two bounded and
continuous derivatives, where $g_{2i}$ is the $i$th component of
$g_{2}(t)$, $1\leq i\leq q$;
  \\
  (ii)\ $g_{3j}$ satisfies a Lipschitz condition of order 1, where $g_{3j}$ is
the $j$th component of $g_{3}(t)$, and  $g_{3}(t)=E(X|X^{\rm
T}\beta_0=t)$, $1\leq j\leq p$. }
 \vskip 10pt

\toto{C3.}{(i)\ The kernel $K$ is a bounded, continuous and
symmetric probability density function, satisfying
$$
\int_{-\infty}^\infty\!u^2K(u)du\neq 0,\ \
\int_{-\infty}^\infty\!|u|^{2}K(u)du<\infty;
$$ }

\toto{\qquad}{(ii)\ $K$ satisfies a Lipschitz condition on $R^1$. }
 \vskip 7pt

\toto{C4.}{(i)\ $\sup_tE\big(\|Z\|^2|X_1^{\rm
T}\beta_0=t\big)<\infty$;
  \\
  (ii)\ $E(e)=0$, ${\rm Var}(e)=\sigma^2<\infty$, $E(e^4)<\infty$. }
 \vskip 10pt

\toto{C5.}{(i)\ $nh^2/\log^2n\rightarrow\infty$,
 ${\displaystyle \limsup_{n\rightarrow\infty}}\,nh^5<\infty$;
  \\
 (ii)\ $nhh_1^3/\log^2n\rightarrow\infty$, $nh^4 \rightarrow
 0$, ${\displaystyle \limsup_{n\rightarrow\infty}}\,nh_1^5<\infty$. }
 \vskip 10pt

\toto{C6.}{(i)\ ${\bf\Sigma}=\mbox{Cov}\big(Z-E(Z|X^{\rm
T}\beta_{0})\big)$ is a positive definite matrix;
  \\
 (ii)\ ${\bf V}=E\big[g'(X^{\rm T}\beta_0)^2{\bf J}_{\beta_0^{(r)}}^{\rm T}
 XX^{\rm T}{\bf J}_{\beta_0^{(r)}}\big]$
is a positive definite matrix, where ${{\bf J}}_{\beta_0^{(r)}}$
is defined by (\ref{(7)}). }
 }

\begin{rema}\rm\
The Lipschitz condition and the two derivatives in {\rm C1}
and {\rm C2} are standard smoothness conditions. {\rm C3} is the
usual assumption for second-order kernels. {\rm C1} is used to bound
the density function of $X^{\rm T}\beta$ away from zero. This
ensures that the denominators of $\hat{g}(t;\beta,\theta_0)$ and
$\hat{g}'(t;\beta,\theta_0)$ are, with high probability, bounded
away from 0 for $t=x^{\rm T}\beta$, $x\in A$ and $\beta$ near
$\beta_0$.
{\rm C4} is a necessary condition for the asymptotic normality of an
estimator. In {\rm C5(i)}, the range of $h$ for the estimators
$\hat{\theta}$ and $\hat{g}$ is fairly large and contains the rate
$n^{-1/5}$ of ``optimal" bandwidths. However, when analyzing the
asymptotic properties of the estimator $\hat \beta$ of $\beta_0$, we
have to estimate the derivative $g'$ of $g$. As is well known, the
convergence rate of the estimator of $g'$ is slower than that  of
the estimator of $g$ if the same bandwidth is used. This leads to a
slower convergence rate for $\hat \beta$ than $\sqrt n$, unless we
use a kernel of order 3 or  {\it  undersmoothing} to deal with the
bias of the estimator. 
This motivates the introduction of another bandwidth
$h_1$ in {\rm C5(ii)} to control the variability of the estimator of
$g'$, and condition {\rm C5(ii)} for bandwidths $h$ and $h_1$. {\rm
Chiou} and {\rm M\"uller (1998)} also consider the use of two
bandwidths to construct the estimator of $\beta$ in a relevant
model. {\rm C6} ensures that the limiting variances for the
estimators
$\hat{\theta}$ and $\hat{\beta}$ exist. 
\end{rema}

The following theorems state the asymptotic behavior of the
estimators proposed in Section 2.1. We first establish the
asymptotic efficiency of $\hat \theta$.

\begin{theo}
\label{theo1}\ Suppose that conditions {\rm C1, C2(i), C3(i),
C4(i), C5(i)} and {\rm C6(i)} hold. When
$\|\hat{\beta}_Z-\beta_Z\|=O_P\big( n^{-1/2}\big)$ and
$\|\hat{\beta}_0-\beta_0\|=O_P\big( n^{-1/2}\big)$, we have
$$
\sqrt{n}(\hat{\theta}-\theta_0)\stackrel{D}{\longrightarrow}N(0,\sigma^2{\bf
\Sigma}^{-1}).
$$
\end{theo}

\begin{rema}\rm\
Carroll et al.(1997)  give  similar results with
$\beta =1$ and $p=1$ (The case of a partially linear model).
Theorem~\ref{theo1} generalizes their Theorems~2 and 3.
\end{rema}

In Theorem~\ref{theo1}, when we start with  $\sqrt{n}$-consistent
estimators for $\beta_Z$ and  $\beta_0$, $\hat{\theta}$ is
consistent for $\theta_0$ with the same asymptotic efficiency as
an estimator that we would have obtained had we known  $\beta_0$
and $g$, and thus the oracle property. Numerous examples of
$\sqrt{n}$- consistent estimators already exist in the literature.
For instance, Hall (1989) showed that one can obtain a
$\sqrt{n}$-consistent estimator for $\beta_0$ using projection
pursuit regression. Under the linearity condition that is slightly
weaker than elliptical symmetry of $X$, Li (1991), Hsing and
Carroll (1992) and Zhu and Ng (1995) proved that SIR,  proposed by
Li (1991), leads to a $\sqrt{n}$-consistent estimator of $\beta_Z$
and of $\beta_0$, the latter  when $Z$ is not present in (1.1). Li
and Zhu (2007) further show that, when including a bias-correction
and under a condition almost equivalent to normality of $X$,
sliced average variance estimation (SAVE, Cook and Weisberg 1991)
performs similarly. We expect the results for $\beta_0$ to hold
when $Z$ is dependent of $X$, provided
 a good estimator of $\beta_Z$ is available.
Under very general regularity conditions and for $q=1$, 
Xia, Tong, Li, and Zhu (2002) proposed the minimum average variance
estimation (MAVE) and  Xia (2006) a refined version of MAVE, and
both methods can provide $\sqrt{n}$-consistent estimators for the
single-index $\beta_0$. However,  there is no result in the
literature regarding  MAVE  when the dimension of $Z$ is larger than 1,  and 
 the $\sqrt{n}$-consistency
needs further study when  $d$ is larger than or equal to $3$, even
for univariate $Z$. Therefore, for general theory, SIR may be a good
choice for the initial estimators of $\beta_Z$ and $\beta_0$.

\begin{theo}\label{theo2}\  Suppose that conditions {\rm C1--C6}
hold. If the $r$th component of $\beta_0$ is positive, we have
$$
\sqrt{n}\big(\hat{\beta}-\beta_0\big)\stackrel{D}{\longrightarrow}
N\big(0,\sigma^2{\bf J}_{\beta_0^{(r)}}{\bf V}^{-1}{\bf Q}{\bf
V}^{-1}{\bf J}_{\beta_0^{(r)}}^{\rm T}\big),
$$
where
 ${\bf Q}=E\big\{g'(X^{\rm T}\beta_0)^2{\bf J}_{\beta_0^{(r)}}^{\rm T}
  [X-E(X|X^{\rm T}\beta_0)][X-E(X|X^{\rm T}\beta_0
  )]^{\rm T}{\bf J}_{\beta_0^{(r)}}\big\}$,
${\bf V}$ and ${\bf J}_{\beta_0^{(r)}}$ are defined in condition C6.
\end{theo}

From H\"ardle et al (1993) and Carroll et al (1997), we can see that
the estimator $\hat \beta$ of $\beta$ has an asymptotic variance
that corresponds to  
a generalized inverse $\sigma^2{\bf Q}_1^-$ where
$${\bf Q}_1=E\left\{g'(X^{\rm T}\beta_0)^2\left[X-E(X|X^{\rm T}\beta_0)\right]
\left[X-E(X|X^{\rm T}\beta_0)\right]^{\rm T}\right\}.$$ Note that
there may be infinitely many inverse matrices of ${\bf Q}_1$, but
there is a unique generalized inverse associated with the Jacobian
$J_{\beta_0^{(r)}}$. The following theorem shows that the
variance-cavariance matrix in Theorem~2 is smaller than
$\sigma^2{\bf Q}_1^-$, the variance associated with ${\bf
J}_{\beta_0^{(r)}}$, in the sense that $\sigma^2{\bf Q}_1^-
-\sigma^2{\bf J}_{\beta_0^{(r)}}{\bf V}^{-1}{\bf Q}{\bf
V}^{-1}{\bf J}_{\beta_0^{(r)}}^{\rm T}$ is a non-negative definite
matrix. We use the usual notation: for two non-negative matrices
${\bf A}$ and ${\bf B}$, ${\bf A}\ge {\bf B}$ denotes that ${\bf
A}-{\bf B}$ is a non-negative definite matrix.

\begin{theo}\label{theo3}\ Under the conditions of Theorem~2, we have

{\rm i)} there is a generalized inverse of ${\bf Q}_1$ that is of
the form ${\bf J}_{\beta_0^{(r)}}^{\rm T}{\bf Q}^{-1}{\bf
J}_{\beta_0^{(r)}}$;

{\rm ii)} ${\bf J}_{\beta_0^{(r)}}^{\rm T}{\bf Q}^{-1}{\bf
J}_{\beta_0^{(r)}} \ge {\bf J}_{\beta_0^{(r)}}{\bf V}^{-1}{\bf
Q}{\bf V}^{-1}{\bf J}_{\beta_0^{(r)}}^{\rm T}.$ \end{theo}

\begin{rema}\rm\ Theorem~\ref{theo3} shows that our
estimator of $\beta_0$ is asymptotically more efficient than those
of H\"ardle et al.(1993) and of Carroll et al. (1997). 
In addition, Carroll et al.(1997) use an iterated procedure to
estimate $\beta_0$ and $\theta_0$,  while our estimation procedure
does not require iteration.\end{rema}

From Theorem~\ref{theo2}, we obtain an asymptotic result regarding
the angle between $\hat \beta$ and $\beta_0$, which can be  used to
study issues of sufficient dimension reduction (SDR).   We refer to
Cook (1998, 2007) for more details.

\begin{coll}\label{coll1}\ Suppose that the conditions of Theorem 2
hold. Then
$$
 \cos(\hat{\beta},\beta_0)-1=O_P\big(n^{-1/2}\big),
$$
where $\cos(\hat{\beta},\beta_0)$ is the cosine of the angle
between $\hat{\beta}$ and $\beta_0$.
\end{coll}

The next two theorems provide the convergence rate of the estimator
$\hat{g}^*(\cdot)$ of $g(\cdot)$ and the asymptotic normality of the
estimator of $\sigma^2$.

\begin{theo}\label{theo4}\ Suppose that the conditions of Theorem 1
hold. If $\|\hat{\beta}-\beta_0\|=O_P\big( n^{-1/2}\big)$. Then
$$
\sup_{(x,\beta)\in{\cal A}_n} \big|\hat{g}^*(x^{\rm
T}\beta)-g(x^{\rm T}\beta_0)\big|=O_P\big((nh/\log n )^{-1/2}\big),
$$
where ${\cal A}_n=\{(x,\beta): (x,\beta)\in A\times R^p,
\|\beta-\beta_0\|\leq cn^{-1/2}\}$  for a constant $c>0$.
\end{theo}

%
%
%

\begin{theo}\label{theo5}\ Suppose that conditions  C1--C6
hold and $0<{\rm Var}(e_1^2)<\infty$. Then
$$
 \sqrt{n}(\hat{\sigma}^2-\sigma^2)/({\rm Var}(e_1^2))^{1/2}\stackrel{D}{\longrightarrow}N(0,1).
$$
\end{theo}

Note that $n^{-1}\tilde{\bf Z}^{\rm T}\tilde{\bf
Z}\stackrel{P}{\longrightarrow}{\bf\Sigma}$ in Lemma A.5 of the
Appendix. By Theorems~\ref{theo1} and ~\ref{theo4} , we  obtain
$$
(\tilde{\bf Z}^{\rm T}\tilde{\bf
Z})^{1/2}(\hat{\theta}-\theta_0)/\hat{\sigma}\stackrel{D}{\longrightarrow}N(0,{\bf
I}_q).
$$

We are now in the position to construct confidence regions for
$\theta_0$. From Theorem 10.2d in Arnold (1981) we obtain the
following result.

\begin{theo}\label{theo6}\ Under the conditions of Theorem~\ref{theo5}, we have
$$
(\hat{\theta}-\theta_0)^{\rm T}(\tilde{\bf Z}^{\rm T}\tilde{\bf
Z})(\hat{\theta}-\theta_0)/\hat{\sigma}^2\stackrel{D}{\longrightarrow}\chi_q^2,
$$
where $\chi_q^2$ is chi-square distributed with $q$ degrees of
freedom. Let $\chi_q^2(1-\alpha)$ be the $(1-\alpha)$-quantile of
$\chi_q^2$ for $0<\alpha<1$, an asymptotic confidence region of
$\theta_0$ is
$$
R_{\alpha}=\{\theta: (\hat{\theta}-\theta)^{\rm T}(\tilde{\bf
Z}^{\rm T}\tilde{\bf Z})(\hat{\theta}-\theta)/\hat{\sigma}^2\le
\chi_q^2(1-\alpha)\}.
$$
\end{theo}

To construct confidence regions for $\beta_0$, a plug-in estimator
of the limiting variance of $\hat \beta$ is needed. We respectively
define the following estimators $\hat{\bf V}$ and $\hat{\bf Q}$ of
$\bf V$ and $\bf Q$ by
$$
 \hat{\bf V}=\frac{1}{n}\sum_{i=1}^n\hat{g}'(X_i^{\rm T}\hat{\beta};\hat{\beta},\hat{\theta})^2
 {\bf J}_{\hat{\beta}^{(r)}}^{\rm T} X_iX_i^{\rm T}{\bf J}_{\hat{\beta}^{(r)}}
$$
and
$$
 \hat{\bf Q}=\frac{1}{n}\sum_{i=1}^n\hat{g}'(X_i^{\rm T}\hat{\beta};\hat{\beta},\hat{\theta})^2
 {\bf J}_{\hat{\beta}^{(r)}}^{\rm T}\big[X_i-\hat{g}_3(X_i^{\rm T}\hat{\beta};\hat{\beta})\big]
 \big[X_i-\hat{g}_3(X_i^{\rm T}\hat{\beta};\hat{\beta})\big]^{\rm T}{\bf J}_{\hat{\beta}^{(r)}},
$$
 where $\hat{g}_3(t;\hat{\beta})=\sum_{i=1}^nW_{ni}(t;\hat{\beta})X_i$
is the estimator of $g_3(t)=E(X|X^{\rm T}\beta_0=t)$ and ${\bf
J}_{\hat{\beta}^{(r)}}$ is the estimator of ${\bf
J}_{\beta_0^{(r)}}$. It is easy to prove that ${\bf
J}_{\hat{\beta}^{(r)}}\stackrel{P}{\longrightarrow}{\bf
J}_{\beta_0^{(r)}}$, $\hat{\bf V}\stackrel{P}{\longrightarrow}{\bf
V}$ and $\hat{\bf Q}\stackrel{P}{\longrightarrow}\bf Q$. Then for
any $p \times l$ matrix ${\bf A}$ of full rank with $l<p$,
Theorems~\ref{theo2} and \ref{theo5} imply that
 $$
  \big(n^{-1}{\bf A}^{\rm T}{\bf J}_{\hat{\beta}^{(r)}}\hat{\bf V}^{-1}\hat{\bf Q}\hat{\bf V}^{-1}{\bf J}_{\hat{\beta}^{(r)}}^{\rm T}{\bf A}\big)^{-1/2}{\bf A}^{\rm T}(\hat{\beta}-\beta_0)/\hat{\sigma}
  \stackrel{D}{\longrightarrow}N(0,{\bf I}_l).
 $$
We again use Theorem 10.2d in Arnold (1981) to obtain the following
limiting distribution.

\begin{theo}\label{theo7}\ Suppose that the conditions
 of Theorem~\ref{theo5}
hold. Then
$$
  (\hat{\beta}-\beta_0)^{\rm T}{\bf A}\big(n^{-1}{\bf A}^{\rm T}{\bf J}_{\hat{\beta}^{(r)}}
  \hat{\bf V}^{-1}\hat{\bf Q}\hat{\bf V}^{-1}
  {\bf J}_{\hat{\beta}^{(r)}}^{\rm T}{\bf A}\big)^{-1}
  {\bf A}^{\rm T}(\hat{\beta}-\beta_0)/\hat{\sigma}^2
  \stackrel{D}{\longrightarrow}\chi_l^2.
 $$
 The asymptotic confidence region of ${\bf A}^{\rm T}\beta_0$ is,
letting $\chi_l^2(1-\alpha)$ be the $(1-\alpha)$-quantile of
$\chi_l^2$ for $0<\alpha<1$,
$$
R_{\alpha}=\{{\bf A}^{\rm T}\beta: (\hat{\beta}-\beta)^{\rm T}{\bf
A}\big(n^{-1}{\bf A}^{\rm T}{\bf J}_{\hat{\beta}^{(r)}}\hat{\bf
V}^{-1}\hat{\bf Q}\hat{\bf V}^{-1}{\bf J}_{\hat{\beta}^{(r)}}^{\rm
T}{\bf A}\big)^{-1}
  {\bf A}^{\rm T}(\hat{\beta}-\beta)/\hat{\sigma}^2\le \chi_l^2(1-\alpha)\}.
$$
\end{theo}

\section{Simulation study }

 \hskip\parindent
 In this section, we examine the performance of the procedures in
Section 2, for the estimation of both $\beta_0$ and $\theta_0$. We
report the accuracy of estimators using  PPR and SIR as
dimension-reduction methods. The sample size for the simulated
data is $n=100$ and the number of simulated samples is $2000$ for
the parametric components. When SIR is applied, using 5 or 10
elements per slice generally yields good results. In other words,
each slice contains 10 to 20 points. A quadratic model of the form
$$
Y = (X^{\rm T}\beta_0-0.5)^2 + Z\theta_0 + 0.2e,
$$
was used, where $\theta_0=1$ is a scalar,  $\beta_0=(0.75, 0.5,
-0.25, -0.25, 0.25)^{\rm T}$, $X$ is a 5-dimensional vector with
independent uniform [0,1] components, and $e$ is a standard normal
variable. The dependency between $X$ and $Z$ was prescribed by
defining $Z$ as  a binary variable with probability $\exp(\beta_Z
X)/(1 + \exp(X^{\rm T}\beta_Z))$ to be $1$ and $0$ otherwise. Two
extreme cases of $\beta_Z$ are reported in Table 1 and Table 2, one
based n choosing  the same value as $\beta_0$ with $\beta_Z =
\beta_0$, and the other on $\beta_Z = (0.5, 0, 0.5, 0.5, -0.5)^{\rm
T}$, so that $\beta_Z$ is orthogonal to $\beta_0$. We  also checked
scenarios where $\beta_Z$ and $\beta_0$ are
 neither orthogonal nor parallel to each other, and the results  are in
 agreement with the two extreme cases reported here.

For the smoothing steps, we used a local linear smoother with a
Gaussian kernel throughout.  A product Gaussian kernel was used
when bivariate smoothing was involved and equal bandwidths were
selected for each kernel to save computing time. A pilot study
revealed that the bandwidth chosen at the first stage to estimate
the residual $\eta$ has little effect on the accuracy of the final
estimates of $\theta_0$, so we choose an initial bandwidth of
$0.5$ to estimate $\phi$ in (\ref{(2)}), as this value was
frequently selected by generalized cross validation (GCV).   The
subsequent smoothing steps utilized the GCV method as proposed in
Craven and Wahba (1979). For instance, when estimating $g$ and
$\theta_0$ in the second stage, the GCV statistic is given by the
formula
\begin{equation}
\mbox{GCV}(h) = \frac{1}{n}\sum_{i = 1}^n (Y_i-Z_i^{\rm
T}\hat{\theta} - \hat{g}_h(X_i^{\rm
T}\hat{\beta};\hat{\beta},\hat{\theta})^2/(n^{-1}\mbox{tr}({\bf I} -
{\bf S}_h))^2,
 \label{(9)}
\end{equation} where $\hat{g}_h(\cdot)$ is the estimator of
$g(\cdot)$ with a bandwidth $h$ and ${\bf S}_h$ is the smoothing
matrix corresponding to a bandwidth of $h$. The GCV bandwidth was
selected to minimize (\ref{(9)}).  We use the optimal bandwidth,
$\hat{h}_{\rm opt}$, for $\hat{g}$ and $\hat{\theta}$. When
calculating the estimator $\hat{\beta}$, we chose the bandwidths,
\begin{equation}
h=\hat{h}_{\rm opt}n^{1/5}n^{-1/3}=\hat{h}_{\rm opt}n^{-2/15}\ \
{\rm and}\ \ \hat{h}_1=\hat{h}_{\rm opt},
 \label{(10)}
\end{equation} respectively, because this guarantees that the required bandwidth
has the correct order of magnitude for optimal asymptotic
performance [see Carroll al.(1997), Stute and Zhu (2005), and Zhu
and Ng (2003)]. Note that choices (\ref{(10)}) satisfy condition
C5(ii). Relevant discussion on choosing two distinct bandwidths can
be found in Chiou and M\"uller (1998).



In the simulation, PPR and SIR were used to obtain the initial
estimators of $\beta_0$ and $\beta_Z$. The notation SIR$_c$ means
that when we used SIR to estimate $\beta_Z$, the number of data
points per slice is $c$. The resulting estimates for $\theta_0$
and the one-step iterated estimates are summarized in Tables 1 and
2, where we report
 bias, standard deviation (SD), and mean square error
(MSE). The case with known $\beta_0$ is also reported in the last
row and serves as a gold standard. The right columns under
``One-step iterated estimate'' in Tables 1 and 2 represent the
results obtained when iterating the algorithms in Section 2.1 one
more time after obtaining   the estimates in the left columns.

\

\begin{center}
Tables 1 and 2 are about here
\end{center}

\

From Tables 1 and 2 we find that the three methods have small mean
square errors with projection pursuit regression outperforming both
SIR procedures. This is expected, as the simulated model structure
satisfies the additive assumption of PPR and the estimates of the
$\beta$-directions were iteratively updated through estimates of the
unknown link functions, $\phi$ and $g$. In other non-additive
situations, SIR might be more reliable than PPR. Iterated estimates
improved the results for all cases and markedly so for the
orthogonal case. Compared to the case when $\beta_0$ is known, PPR
typically  attains  80\% or more of the efficiency after one
iteration.

For the estimation of $\beta_0$, we computed the angle (in radians)
between $\hat{\beta}$ and $\beta_0$  as a measure of accuracy.
 The mean, standard deviation (SD),
and  mean squared error (MSE) of the angle between $\hat{\beta}$ and
$\beta_0$ are reported in  Table 3. Here, PPR leads to by far
superior estimates compared to SIR.

\begin{center}
Table 3 is about here
\end{center}

\

The performance of the nonparametric estimates for $g$ is
demonstrated in Figure 1. Again, GCV was used for bandwidth choice
and compared to the estimates based on the optimal fixed bandwidth.
The true function $g$ and the mean of each estimated $g$-function
over the 2000 replicates are plotted. In general, GCV seems to work
well for all parametric and nonparametric components. This is
consistent with the results reported in Chen and Shiau (1994) for
the analysis of partially linear models based on generalized cross
validation (GCV). Theoretical properties of the current models in
regard to  GCV will be  a topic for further investigation.

\

\begin{center}
Figure 1 is about here
\end{center}

\


A final remark is that we tried to compare  our procedure with that
proposed in Carroll, {\it et al.} (1997), for the quadratic model
used in the above simulations with $\beta_Z$ and $\beta_0$
orthogonal. However, we were not able to obtain any results for the
method in Carroll et al. (1997), as their procedure seems to be very
sensitive to the choice of the initial estimates.  We then used our
estimates for $\beta_0$ and $\theta_0$ as the initial values for
their procedure. Nevertheless, we were still unable to obtain any
meaningful comparison results as out of the seven attempted trials
their procedure crashed six times on the first simulation and once
on the second simulation. Since $\theta_0$ is only a scalar, we
postulate that their procedure has difficulties with high
dimensional $\beta_0$, which is here a five-dimensional vector.

\section{Data Example }

 \hskip\parindent
We  analyze the Boston Housing data mentioned in Section 1.   The
goal is  to determine the effect of the various  variables on
housing price, including a binary variable, which describes whether
the census tract borders the Charles River. According to Harrison
and Rubinfeld (1978), bordering the river should have a positive
effect on the median housing price of the census tract. They used a
linear model that included a log transformation for the response
variable and three of the covariates, and power transformations for
three other covariates.  Their final model  is
\begin{eqnarray*}
\log(MV) & = & a_1 + a_2 RM^2 + a_3 AGE + a_4 \log(DIS) + a_5 \log(RAD) + a_6 TAX \\
&& \mbox{} + a_7 PTRATIO + a_8 (B - 0.63)^2 + a_9 \log(LSTAT) +
a_{10}
CRIM \\
&& \mbox{} + a_{11} ZN + a_{12} INDUS + a_{13} CHAS + a_{14} NOX^p +
e.
\end{eqnarray*}
The coefficient $a_{13}$ is estimated to be 0.088, which is
significant with a $p$-value of less than $0.01$ for the hypothesis
$H_0: a_{13} = 0$ versus $H_1: a_{13} \neq 0$.  The coefficient of
determination $R^2$ attained by their analysis is $0.81$, where
$R^2$ is the squared correlation between the true
dimension-reduction variable $X^{\rm T}\beta_0$ and the estimated
dimension-reduction variable $X^{\rm T}\hat{\beta}_0$.

This data set was also analyzed by Chen and Li (1998), who used
sliced inverse regression with all thirteen covariates. After
examining the initial results, Chen and Li (1998) trimmed the data
and then dropped some of the variables. We fit the data on the first
SIR direction of the initial analysis reported in their article and
obtained an $R^2$ of $0.705$ using GCV bandwidth  $0.43$. Note that
the assumptions of sliced inverse regression are probably not met
because some of the covariates are discrete. We thus proposed to use
a partial-linear single-index model. Several choices of $Z$ were
attempted, but they did not yield better results, in terms of $R^2$,
than the one using only the Charles River variable as $Z$ and the
other covariates as $X$. We thus focus on this model, where a log
transformation was applied on $Y$.

To select the number of observations per slice in the dimension
reduction step of SIR, we borrow our experience in the simulation
presented in Section 3, where 5 or 10 observations per slice
worked well for a total sample size of 100, leading to about 20 to
10 slices. Since the sample size for the housing data is much
larger, we use SIR with 20 data points per slice and this leads to
a total of 26 slices. As Chen and Li (1998) pointed out, SIR is
not sensitive to the choice of slice number, and they tried
slicing with 10 or 30 points per slice leading to 17 or 50 slices,
and obtain very similar results. The GCV bandwidth for estimating
$g$ and $\theta$ is 0.367, which is smaller than the bandwidth
0.43 chosen by the GCV method for the SIR approach of Chen and Li
(1998). To estimate $\beta$ by (2.10), the bandwidths selected by
(3.2) for $h=0.16$ and for $h_1$ is 0.367. The $R^2$ is 0.8047,
which is essentially equal to that obtained by Harrison and
Rubinfeld and higher than that using SIR on all thirteen
variables. The value of the test statistic for $H_0: \theta = 0$
versus $H_1: \theta \neq 0$ is 3.389 when the degrees of freedom
are calculated according to Hastie and Tibshirani, and 3.419 when
$n$ degrees of freedom are used. Either way the result is
significant with $p$-value $< 0.01$.

We also omitted the Charles River variable and used  a
dimension-reduction model on $Y$ and $X$. After obtaining an
estimate for $\beta_0$, we then estimate the relationship between
$Y$ and $X^{\rm T}\hat{\beta}$.  GCV yields a bandwidth of 0.16, and
we obtain $R^2 = 0.8021$.  Even though the Charles River variable is
significant, its inclusion leads to only a minor increase in $R^2$.

\

\begin{center}
Figure 2 is about here
\end{center}

\

Figure 2 shows the estimated $g$ along with the data. On the
$x$-axis of the above graph, the estimated value $x^{\rm
T}\hat{\beta}$ is given, and on the $y$-axis, the estimated value
$\hat{g}^*(t)$. Figure 2 shows a downward trend in the effective
dimension reduction (EDR) variate obtained. The upward curvature of
the function at high values of the EDR variate may or may not be a
real effect.

The advantage of our procedure over the one used by Harrison and
Rubinfeld is that Harrison and Rubinfeld have to make choices
regarding transformations for every variable in the model. We  only
need to choose the bandwidth or bandwidths used for smoothing.

\section{Proofs of Theorems }

\hskip\parindent  Since the proofs of the theorems are rather long,
the proofs of Theorems 1--4 are presented in this section, and more
details of the proofs are divided into Lemmas A.2--A.7 in the
Appendix.

In this section and the Appendix, we use $c>0$ to represent any
constant which may take different values for each appearance, and
$a\wedge b=\min(a,b)$.

\

{\bf Proof of Theorem~\ref{theo1}.}\ \ Denote
$$
\tilde{G} =\big(g(X_1^{\rm T}\beta_0)-\hat{g}(X_1^{\rm
T}\hat{\beta}_0;\hat{\beta}_0,\theta_0),\ldots,g(X_n^{\rm
T}\beta_0)-\hat{g}(X_n^{\rm
T}\hat{\beta}_0;\hat{\beta}_0,\theta_0)\big)^{\rm T}.
$$
From (\ref{(5)}) we have
$$\align
\sqrt{n}(\hat{\theta}-\theta_0)
 = & \sqrt{n}(\tilde{\bf Z}^{\rm T}\tilde{ \bf Z})^{-1}\tilde{\bf Z}^{\rm T}\tilde{G}
  + \sqrt{n}(\tilde{\bf Z}^{\rm T}\tilde{ \bf Z})^{-1}\tilde{\bf Z}^{\rm T} e.
\endalign $$
Lemma A.5 in the Appendix implies
\begin{equation}
n(\tilde{\bf Z}^{\rm T}\tilde{\bf
Z})^{-1}\stackrel{P}{\longrightarrow}\Sigmavec^{-1}.
 \label{(11)}
\end{equation} Therefore, Lemma A.6 in the Appendix leads to
$$
\sqrt{n}(\tilde{\bf Z}^{\rm T}\tilde{ \bf Z})^{-1}\tilde{\bf Z}^{\rm
T}\tilde{G} \stackrel{P}{\longrightarrow}0.
$$
It remains to show that
\begin{equation}
\sqrt{n}(\tilde{\bf Z}^{\rm T}\tilde{\bf Z})^{-1}\tilde{\bf Z}^{\rm
T} e \stackrel{D}{\longrightarrow}N(0,\sigma^2{\bf\Sigma}^{-1}).
 \label{(12)}
\end{equation}
Since
$$\align
\tilde{\bf Z}^{\rm T} e
 & = \sum_{i=1}^n\big[Z_i-g_2(X_i^{\rm T}\beta_0)\big]e_i
   + \sum_{i=1}^n\big[g_2(X_i^{\rm T}\beta_0)-\hat{g}_2(X_i^{\rm T}\hat{\beta}_0;\hat{\beta}_0)\big]e_i
 \\
 & =: M_1+M_2.
\endalign $$
The central limit theorem implies
 $n^{-1/2}M_1\stackrel{D}{\longrightarrow}N(0,{\bf \Sigma})$.
Similarly to the proof of (A.17), it is easy to obtain that
 $n^{-1/2}M_2\stackrel{P}{\longrightarrow}0$.
This together with (\ref{(11)}) and Slutsky's Theorem proves
(\ref{(12)}), and hence Theorem 1. \hfill $\fbox{}$

\

{\bf Proof of Theorem~\ref{theo2}.}\ \ The proof is divided into two
steps:  From (\ref{(8)}), step (I) provides the existence of the
least squares estimator $\hat{\beta}$ of $\beta_0$, and from
(\ref{(9)}), step (II) proves the asymptotic normality of
$\hat{\beta}$.

(I)\ \ \textbf{Proof of existence.} We prove the following fact:
Under conditions C1--C5 and with probability one there exists an
estimator  of $\beta_0$ minimizing expression (\ref{(8)}) in ${\cal
B}_{1n}$, where ${\cal B}_{1n}=\big\{\beta:
\|\beta-\beta_0\|=B_1n^{-1/2}\big\}$ for some constant such that
$0<B_1<\infty$.

In fact, let ${\bf Y}=(Y_1,\ldots,Y_n)^{\rm T}$ and ${\bf
Z}=(Z_1,\ldots,Z_n)^{\rm T}$. We have
$$\align
D(\beta)
 & = ({\bf Y}-{\bf Z}\hat{\theta})^{\rm T}({\bf I}-{\bf S}_\beta)^{\rm T}({\bf I}-{\bf S}_\beta)({\bf Y}-{\bf
 Z}\hat{\theta})\\
 & = ({\bf Y}-{\bf Z}\theta_0)^{\rm T}({\bf I}-{\bf S}_\beta)^{\rm T}({\bf I}-{\bf S}_\beta)({\bf Y}-{\bf
 Z}\theta_0) \\
 & \quad~
  - 2({\bf Y}-{\bf Z}\theta_0)^{\rm T}({\bf I}-{\bf S}_\beta)^{\rm T}({\bf I}-{\bf S}_\beta){\bf Z}(\hat{\theta}-\theta_0) \\
 & \quad~ + \{{\bf Z}(\hat{\theta}-\theta_0)\}^{\rm T}({\bf I}-{\bf S}_\beta)^{\rm T}({\bf I}-{\bf S}_\beta){\bf Z}(\hat{\theta}-\theta_0) \\
 & =: D_1(\beta)-D_2(\beta) + D_3(\beta).
\endalign $$
The same arguments as in the proof of Theorem 1 can be used to
obtain that
 $D_2(\beta)=R_0+o_P(1)$ and $D_3(\beta)=o_P(1)$, where $R_0$ is a constant independent of $\beta$. This implies
 $D(\beta)=D_1(\beta)-R_0+o_P(1)$. Thus, minimizing $D(\beta)$ simultaneously with respect to $\beta$ is
very much like separately minimizing $D_1(\beta)$ with respect to
$\beta$. It follows from (\ref{(6)}) that we only need to prove the
existence of an estimator of $\beta_0^{(r)}$ in ${\cal B}_{2n}$,
where ${\cal B}_{2n}=\big\{\beta^{(r)}:
\|\beta^{(r)}-\beta_0^{(r)}\|=B_2n^{-1/2}\big\}$ for some constant
such that $0<B_2<\infty$. Since
$R(\beta^{(r)})=(-\frac{1}{2})\frac{\partial
D_1(\beta)}{\partial\beta^{(r)}}$, where $R(\beta^{(r)})$ is defined
in (A.19) of Lemma A.7. For an arbitrary $\beta^{(r)}\in {\cal
B}_{2n}$ with the value of constant $B_2$ in ${\cal B}_{2n}$ to be
determined, we have from Lemma A.7 below that
\begin{eqnarray}
&&(\beta^{(r)}-\beta_0^{(r)})^{\rm T} R(\beta^{(r)})\nonumber \\
&& \quad = (\beta^{(r)}-\beta_0^{(r)})^{\rm T} U(\beta_0^{(r)})
   -n(\beta^{(r)}-\beta_0^{(r)})^{\rm T}{\bf V}(\beta^{(r)}-\beta_0^{(r)})
   +o_P(1).
  \label{(13)}
\end{eqnarray}

The following arguments are similar to those used by Weisberg and
Welsh (1994), which in turn use (6.3.4) of Ortega and Rheinboldt
(1973). We note that term (\ref{(13)}) is dominated by the term
$\sim B_2^2$ because $\sqrt{n}\|\beta^{(r)}-\beta_0^{(r)}\|=B_2$,
whereas $|(\beta^{(r)}-\beta_0^{(r)})^{\rm T}
U(\beta_0^{(r)})|=B_2O_P(1)$ and
$n(\beta^{(r)}-\beta_0^{(r)})^{\rm T}{\bf
V}(\beta^{(r)}-\beta_0^{(r)})\sim B_2^2$. So, for any given
$\eta>0$, if $B_2$ is chosen large enough, then it will follows
that $(\beta^{(r)}-\beta_0^{(r)})^{\rm T} R(\beta_0^{(r)})<0$ on
an event with probability $1-\eta$. From the arbitrariness of
$\eta$, we can prove the existence of the least squares estimator
of $\beta_0^{(r)}$ in ${\cal B}_{2n}$ as in the proof of Theorem
5.1 of Welsh (1989). The details are omitted.

(II)\ \ \textbf{Proof of asymptotic normality.} From step (I) we
find that $\hat{\beta}^{(r)}$ is a solution in ${\cal B}_{2n}$ to
the equation
 $R(\beta^{(r)})=0$. That is, $R(\hat{\beta}^{(r)})=0$.
By Lemma A.7, we have
$$
 0 = U(\beta_0^{(r)})-n{\bf V}(\hat{\beta}^{(r)}-\beta_0^{(r)})+o_P\big(\sqrt{n}\
 \big),
$$
and hence
$$
 \sqrt{n}(\hat{\beta}^{(r)}-\beta_0^{(r)})
 ={\bf V}^{-1}n^{-1/2}U(\beta_0)+o_P(1).
$$
We now consider the estimator $\hat{\beta}$. A simple calculation
yields
$$
\frac{2\sqrt{1-\|\beta_0^{(r)}\|^2}}{\sqrt{1-\|\hat{\beta}^{(r)}\|^2}+\sqrt{1-\|\beta_0^{(r)}\|^2}}-1
=\frac{\sqrt{1-\|\beta_0^{(r)}\|^2}-\sqrt{1-\|\hat{\beta}^{(r)}\|^2}}{\sqrt{1-\|\hat{\beta}^{(r)}\|^2}+\sqrt{1-\|\beta_0^{(r)}\|^2}}
=O_P(n^{-1/2}),
$$
and hence
$$\align
 &\sqrt{1-\|\hat{\beta}^{(r)}\|^2}-\sqrt{1-\|\beta_0^{(r)}\|^2}
 \\
 & \quad~ = -\frac{(\hat{\beta}^{(r)}+\beta_0^{(r)})^{\rm T}(\hat{\beta}^{(r)}-\beta_0^{(r)})}{\sqrt{1-\|\hat{\beta}^{(r)}\|^2}+\sqrt{1-\|\beta_0^{(r)}\|^2}}
 \\
 & \quad~ = -\frac{2\beta_0^{(r)T}(\hat{\beta}^{(r)}-\beta_0^{(r)}) + \|\hat{\beta}^{(r)}-\beta_0^{(r)}\|^2}
 {\sqrt{1-\|\hat{\beta}^{(r)}\|^2}+\sqrt{1-\|\beta_0^{(r)}\|^2}}
  \\
 & \quad~ = -\frac{\beta_0^{(r)T}(\hat{\beta}^{(r)}-\beta_0^{(r)})}{\sqrt{1-\|\beta_0^{(r)}\|^2}}
  + O_P(n^{-1}).
\endalign $$
It follows from (\ref{(6)}) and the above equation, that
\begin{eqnarray*}
\hat{\beta}-\beta_0
 &=& \left(
     \begin{array}{c}
          \hat{\beta}_1 \\
          \vdots \\
          \hat{\beta}_{r-1} \\
          \sqrt{1-\|\hat{\beta}^{(r)}\|^2} \\
          \hat{\beta}_{r+1} \\
          \vdots \\
          \hat{\beta}_p
        \end{array}
      \right)
  - \left(
    \begin{array}{c}
          \beta_{01} \\
          \vdots \\
          \beta_{0(r-1)} \\
          \sqrt{1-\|\beta_0^{(r)}\|^2} \\
          \beta_{0(r+1)} \\
          \vdots \\
          \beta_{0p}
        \end{array}
      \right)
  = \left(
     \begin{array}{c}
          \hat{\beta}_1 - \beta_{01} \\
          \vdots \\
          \hat{\beta}_{r-1} - \beta_{0(r-1)} \\
          -\frac{\beta_0^{(r)T}(\hat{\beta}^{(r)}-\beta_0^{(r)})}{\sqrt{1-\|\beta_0^{(r)}\|^2}} \\
          \hat{\beta}_{r+1} - \beta_{0(r+1)} \\
          \vdots \\
          \hat{\beta}_p - \beta_{0p}
        \end{array}
      \right)
  + O_P(n^{-1}).
\end{eqnarray*}
That is, from the definition of ${\bf J}_{\beta_0^{(r)}}$ of
(\ref{(7)})
$$
\hat{\beta}-\beta_0={\bf
J}_{\beta_0^{(r)}}\big(\hat{\beta}^{(r)}-\beta_0^{(r)}\big)+O_P\big(n^{-1}\big).
$$
Thus, we have
$$
 \sqrt{n}(\hat{\beta}-\beta_0)
 ={\bf J}_{\beta_0^{(r)}}{\bf V}^{-1}n^{-1/2}U(\beta_0^{(r)})+o_P(1).
$$
Theorem 2  follows from this, Central Limit Theorem and Slutsky's
Theorem. \hfill $\fbox{}$

{\bf Proof of Theorem~\ref{theo3}.}\ \  Recalling the definition
of ${\bf Q}$, we can see that ${\bf Q}={\bf
J}_{\beta_0^{(r)}}^{\rm T}{\bf Q}_1{\bf J}_{\beta_0^{(r)}}.$
Define
$${\bf\Pi}_0:={\bf J}_{\beta_0^{(r)}}{\bf Q}^{-1}{\bf J}_{\beta_0^{(r)}}^{\rm T}, \quad
{\bf\Pi}_1:={\bf J}_{\beta_0^{(r)}}{\bf V}^{-1}{\bf Q}{\bf
V}^{-1}{\bf J}_{\beta_0^{(r)}}^{\rm T}.$$ We now prove that
${\bf\Pi}_0$ is a generalized inverse of ${\bf Q}_1$. To this end,
we need to prove that ${\bf\Pi}_0{\bf Q}_1{\bf\Pi}_0={\bf\Pi}_0$
and ${\bf Q}_1{\bf\Pi}_0{\bf Q}_1={\bf Q}_1$. Note that
$${\bf\Pi}_0{\bf Q}_1{\bf\Pi}_0={\bf J}_{\beta_0^{(r)}}{\bf Q}^{-1}{\bf
J}_{\beta_0^{(r)}}^{\rm T}{\bf Q}_1{\bf J}_{\beta_0^{(r)}}{\bf
Q}^{-1}{\bf J}_{\beta_0^{(r)}}^{\rm T}={\bf J}_{\beta_0^{(r)}}{\bf
Q}^{-1}{\bf Q}{\bf Q}^{-1}{\bf J}_{\beta_0^{(r)}}^{\rm
T}={\bf\Pi}_0.$$ We now prove ${\bf Q}_1{\bf\Pi}_0{\bf Q}_1={\bf
Q}_1$. First, by ${\bf Q}{\bf R}$ decomposition (see, e.g.  Gentle
1998, Section 3.2.2, pages 95-97 for more details)  for $\beta_0$,
we can find its orthogonal complement such that ${\bf B}=(b_1,
\beta_0)$ is an orthogonal matrix, and $\beta_0={\bf B}\left(
                                              \begin{array}{c}
                                                {\bf 0} \\
                                                1 \\
                                              \end{array}
                                            \right).$
 Thus, ${\bf
J}_{\beta_0^{(r)}}={\bf B}{\bf B}^{\rm T}{\bf
J}_{\beta_0^{(r)}}=:{\bf B}{\bf R}$ where ${\bf R}=\left(
                                              \begin{array}{c}
                                                {\bf R}_1 \\
                                                {\bf R}_2 \\
                                              \end{array}
                                            \right)
$ with ${\bf R}_1$ being a $(p-1)\times (p-1)$ nonsigular matrix.
Further, note that
\begin{eqnarray*}
{\bf Q}&=&{\bf J}_{\beta_0^{(r)}}^{\rm T}{\bf Q}_1{\bf J}_{\beta_0^{(r)}}={\bf R}^{\rm T}{\bf B}^{\rm T}{\bf Q}_1{\bf B}{\bf R}\\
&=&{\bf R}^{\rm T}\left(
        \begin{array}{cc}
          b_1^{\rm T}{\bf Q}_1b_1 & 0 \\
          0 & 0 \\
        \end{array}
      \right){\bf R}\\
      &=&{\bf R}_1^Tb_1^{\rm T}{\bf Q}_1b_1{\bf R}_1.
\end{eqnarray*}
To prove the result, we rewrite ${\bf Q}_1$ in another form.
Define ${\bf S}={\bf B}\left(
        \begin{array}{cc}
          {\bf R}_1 & 0 \\
          0 & 1 \\
        \end{array}
      \right).$
${\bf S}$ is a nonsingular matrix. Then

\begin{eqnarray*}
{\bf Q}_1&=&({\bf S}^{\rm T})^{-1}{\bf S}^{\rm T}{\bf Q}_1{\bf
S}{\bf S}^{-1}=({\bf S}^{\rm T})^{-1}\left(
        \begin{array}{cc}
          {\bf R}_1^{\rm T} & 0 \\
          0 & 1 \\
        \end{array}
      \right){\bf B}^{\rm T}{\bf Q}_1{\bf B}\left(
        \begin{array}{cc}
          {\bf R}_1 & 0 \\
          0 & 1 \\
        \end{array}
      \right){\bf S}^{-1}\\
      &=&({\bf S}^{\rm T})^{-1}{\bf S}^{\rm T}{\bf Q}_1{\bf S}{\bf S}^{-1}=({\bf S}^{\rm T})^{-1}\left(
        \begin{array}{cc}
          {\bf R}_1^{\rm T} & 0 \\
          0 & 1 \\
        \end{array}
      \right)
\left(
        \begin{array}{cc}
          b_1^{\rm T}{\bf Q}_1b_1 & 0 \\
          0 & 0 \\
        \end{array}
      \right)
      \left(
        \begin{array}{cc}
          {\bf R}_1 & 0 \\
          0 & 1 \\
        \end{array}
      \right){\bf S}^{-1}\\
      &=&({\bf S}^{\rm T})^{-1}\left(
        \begin{array}{cc}
          {\bf Q} & 0 \\
          0 & 0 \\
        \end{array}
      \right){\bf S}^{-1}.
\end{eqnarray*}
We now prove that ${\bf Q}_1{\bf\Pi}_0{\bf Q}_1={\bf Q}_1$ that is
of the above form. From the above and noting that ${\bf
S}^{-1}=\left(
        \begin{array}{cc}
          {\bf R}_1^{-1} & 0 \\
          0 & 1 \\
        \end{array}
      \right){\bf B}^{\rm T}$ and $({\bf S}^{\rm T})^{-1}={\bf B}\left(
        \begin{array}{cc}
          ({\bf R}_1^{\rm T})^{-1} & 0 \\
          0 & 1 \\
        \end{array}
      \right)$, we have
\begin{eqnarray*}
&&{\bf Q}_1{\bf\Pi}_0{\bf Q}_1\\
&=& ({\bf S}^{\rm T})^{-1}\left(
        \begin{array}{cc}
          {\bf Q} & 0 \\
          0 & 0 \\
        \end{array}
      \right){\bf S}^{-1}{\bf B}{\bf R}Q^{-1}{\bf R}^{\rm T}{\bf B}^{\rm T}({\bf S}^{\rm T})^{-1}
\left(
        \begin{array}{cc}
          {\bf Q} & 0 \\
          0 & 0 \\
        \end{array}
      \right){\bf S}^{-1}\\
      &=& ({\bf S}^{\rm T})^{-1}\left(
        \begin{array}{cc}
          {\bf Q} & 0 \\
          0 & 0 \\
        \end{array}
      \right)\left(
        \begin{array}{cc}
          {\bf R}_1^{-1} & 0 \\
          0 & 1 \\
        \end{array}
      \right){\bf B}^{\rm T}{\bf B}{\bf R}{\bf Q}^{-1}{\bf R}^{\rm T}{\bf B}^{\rm T}{\bf B}\left(
        \begin{array}{cc}
          ({\bf R}_1^{\rm T})^{-1} & 0 \\
          0 & 1 \\
        \end{array}
      \right)
\left(
        \begin{array}{cc}
          {\bf Q} & 0 \\
          0 & 0 \\
        \end{array}
      \right){\bf S}^{-1}\\
&=& ({\bf S}^{\rm T})^{-1}\left(
        \begin{array}{cc}
          {\bf Q} & 0 \\
          0 & 0 \\
        \end{array}
      \right)\left(
        \begin{array}{c}
          {\bf 1}  \\
           {\bf R}_2 \\
        \end{array}
      \right){\bf Q}^{-1}\left(
        \begin{array}{cc}
           {\bf 1} & {\bf R}_2
        \end{array}
      \right)
\left(
        \begin{array}{cc}
          {\bf Q} & 0 \\
          0 & 0 \\
        \end{array}
      \right){\bf S}^{-1}\\
&=&({\bf S}^{\rm T})^{-1}\left(
        \begin{array}{c}
          {\bf Q} \\
          0  \\
        \end{array}
      \right){\bf Q}^{-1}\left(
        \begin{array}{cc}
           {\bf Q} & 0
        \end{array}
      \right)
{\bf S}^{-1}=({\bf S}^{\rm T})^{-1}\left(
        \begin{array}{cc}
          {\bf Q} & 0 \\
          0 & 0 \\
        \end{array}
      \right){\bf S}^{-1}={\bf Q}_1.
\end{eqnarray*}
Thus, ${\bf\Pi}_0$ is one of the solutions of ${\bf Q}_1^-$. To
prove that the asymptotic variance-covariance matrix
$\sigma^2{\bf\Pi}_1$ of our estimator is smaller than the
corresponding matrix $\sigma^2{\bf\Pi}_0$ given in  H\"ardle et
al. (1993), we only need to show that ${\bf\Pi}_0- {\bf\Pi}_1$ is
a positive semi-definite matrix, that is, ${\bf\Pi}_0>
{\bf\Pi}_1$. Recall that ${\bf V}={\bf
J}_{\beta_0^{(r)}}^TE\left\{g'(X^{\rm T}\beta_0)^2X X^{\rm
T}\right\}{\bf J}_{\beta_0^{(r)}}$. Note that both ${\bf Q}$ and
${\bf V}$ are positive definite matrices and obviously ${\bf
V}\geq {\bf Q}$. Thus, ${\bf Q}^{-1}\geq {\bf V}^{-1}$, and then
${\bf V}^{-1}\geq {\bf V}^{-1}{\bf Q}{\bf V}^{-1}$. From these two
inequalities, it is easy to see that
$${\bf\Pi}_0\geq {\bf J}_{\beta_0^{(r)}}{\bf V}^{-1}{\bf
J}_{\beta_0^{(r)}}^{\rm T} \geq {\bf J}_{\beta_0^{(r)}}{\bf
V}^{-1}{\bf Q}{\bf V}^{-1}{\bf J}_{\beta_0^{(r)}}^{\rm
T}={\bf\Pi}_1.$$ The proof is now complete. \hfill $\fbox{}$

{\bf Proof of Corollary~\ref{coll1}.}\ \ Let $\bullet$ denote the
inner product of two vectors. Theorem 2 implies  $
\|\hat{\beta}-\beta_{0}\|=O_P\big(n^{-1/2}\big) $ and
$$\align
|\cos(\hat{\beta},\beta_{0})-1|
 & = \big|(\hat{\beta}-\beta_{0})\bullet\beta_{0}\big/\|\hat{\beta}\|
    + (\|\beta_{0}\|-\|\hat{\beta}\|)\big/\|\hat{\beta}\|\,\big| \\
 & \leq 3\|\hat{\beta}-\beta_{0}\|\big/\|\hat{\beta}\|
 =O_P\big(n^{-1/2}\big).
\endalign $$
 This completes the proof of Corollary 1.       \hfill $\fbox{}$

{\bf Proof of Theorem~\ref{theo4}.}\ \ Denote
$\theta_0=(\theta_{01},\ldots,\theta_{0q})^{\rm T}$,
$\hat{\theta}=(\hat{\theta}_1,\ldots,\hat{\theta}_q)^{\rm T}$.
Theorem 1 and Lemma A.4 in the Appendix yield
$$\align
 & \sup_{(x,\beta)\in{\cal A}_n}\big|\hat{g}^*(x^{\rm T}\beta)-g(x^{\rm T}\beta_{0})\big|\\
 &\qquad\leq \sum_{s=1}^q\sup_{(x,\beta)\in{\cal A}_n}\big|\hat{g}_{2s}(x^{\rm T}\beta;\beta)-g_{2s}(x^{\rm T}\beta_0)\big|
   |\hat{\theta}_s-\theta_{0s}|\\
 &\qquad\ \ \ \ +  \sup_{(x,\beta)\in{\cal A}_n}
  \big|\hat{g}(x^{\rm T}\beta;\beta,\theta_0)-g(x^{\rm T}\beta_0)\big| \\
 &\qquad\ \ \ \ + \sum_{s=1}^q\sup_{x\in A}|g_{2s}(x^{\rm T}\beta_0)||\hat{\theta}_s-\theta_{0s}|
 = O_P\big((nh/\log n)^{-1/2}\big),
\endalign $$
and hence Theorem 4 follows.  \hfill $\fbox{}$

{\bf Proof of Theorem~\ref{theo5}.}\ \ Decomposing $\hat{\sigma}^2$
into several parts, we have
$$\align
\hat{\sigma}^2
 & = \frac{1}{n}\sum_{i=1}^ne_i^2 +\frac{1}{n}\sum_{i=1}^n\big[Z_i^{\rm T}(\theta_0-\hat{\theta})
    + g(X_i^{\rm T}\beta_0)-\hat{g}(X_i^{\rm T}\hat{\beta};\hat{\beta},\theta_0)\big]^2\\
  & \quad\, + \frac{2}{n}\sum_{i=1}^ne_iZ_i^{\rm T}(\theta_0-\hat{\theta})
    + \frac{2}{n}\sum_{i=1}^ne_i\big[g(X_i^{\rm T}\beta_0)-\hat{g}(X_i^{\rm T}\hat{\beta};\hat{\beta},\theta_0)\big]\\
 & =: I_1+I_2+I_3+I_4.
\endalign $$
Note that $\sqrt{n}\|\hat{\theta}-\theta_0\|=O_P(1)$ and using (A.8)
of Lemma A.4, we have
$$\align
 \sqrt{n}|I_2|
 & \leq \frac{1}{n}\sum_{i=1}^n\|Z_i\|^2\sqrt{n}\|\hat{\theta}-\theta_0\|^2 \\
 & \quad\, +\sqrt{n}\sup_{(x,\beta)\in {\cal A}_n}|g(x^{\rm T}\beta_0)-\hat{g}(x^{\rm T}\beta;\beta,\theta_0)|^2 \\
 & = O_P\big(n^{-1/2}\big)+O_P\big((nh^2/\log^2n)^{-1/2}\big)\stackrel{P}{\longrightarrow}0.
\endalign $$
Since $Ee_i=0$, we obtain
$\sqrt{n}I_3\stackrel{P}{\longrightarrow}0$. Similarly to the proof
of (A.17) in the Appendix, we also have
$\sqrt{n}I_4\stackrel{P}{\longrightarrow}0$. This proves that
$\hat{\sigma}^2=\frac{1}{n}\sum_{i=1}^ne_i^2+o_P\big(n^{-1/2}\big)$.
Therefore, we have
$$
\sqrt{n}(\hat{\sigma}^2-\sigma^2)=\frac{1}{\sqrt{n}}\sum_{i=1}^n(e_i^2-\sigma^2)+o_P(1).
$$
The proof can now be completed by employing the central limit
theorem. \hfill $\fbox{}$

\

\begin{center}
\bf APPENDIX
\end{center}

The following Lemmas A.1--A.7 are needed to prove Theorems 1, 2, 4,
5. Lemma A.1 gives an important probability inequality and Lemmas
A.2 and A.3 provide bounds for the moments of the relevant
estimators. They are used to obtain the rates of convergence for the
estimators of the nonparametric component, and are used in the proof
of Lemmas A.4--A.7. Lemma A.4 presents the uniform rates of
convergence in probability for the estimators $\hat{g}$, $g_{2s}$
and $\hat{g}'$. These results are very useful for the nonparametric
estimations. The proof of Lemma A.5--A.7, as well as Theorem 3 and
4, rely on Lemma A.4. To simplify the proof of Theorem 1, we divide
the main steps of the proofs into Lemmas A.5 and A.6. Lemma A.5 is
used to obtain the limiting variance of the estimator
$\hat{\theta}$,  and Lemma A.6 together with Lemma A.5 shows  that
the rate of convergence of the nonlinear section of $e_i$ for
$\hat{\theta}-\theta_0$ is $o_P\big(n^{-1/2}\big)$. Lemma A.7
provides  the main step for the proof of Theorem 2.

{\bf Lemma A.1}\ \ {\it Let $\xi_1(x,\beta),\ldots,\xi_n(x,\beta)$
be a sequence of random variables. Denote
$f_{x,\beta}(V_i)=\xi_i(x,\beta)$ for $i=1,\ldots,n$, where
$V_1,\ldots,V_n$ be a sequence of random variables, and
$f_{x,\beta}$ is a function on ${\cal A}_n$, where ${\cal
A}_n=\{(x,\beta): (x,\beta)\in A\times R^p, \|\beta-\beta_0\|\leq
cn^{-1/2}\}$ for a constant $c>0$. Assume that $f_{x,\beta}$
satisfies }
$$
\frac{1}{n}\sum_{i=1}^n\big|f_{x,\beta}(V_i)-f_{x^*,\beta^*}(V_i)|
\leq c n^a\big[\|\beta-\beta^*\|+\|x-x^*\|\big]
  \eqno{\rm(A.1)}
$$
{\it for some constants $x^*$, $\beta^*$, $a>0$ and $c>0$. Let
$\varepsilon_n>0$ depend only on $n$.  If }
$$
P\left\{\bigg|\frac{1}{n}\sum_{i=1}^n\xi_i(x,\beta)\bigg|>\frac{1}{2}\varepsilon_n\right\}\leq
\frac{1}{2},
  \eqno{\rm(A.2)}
$$
{\it for $(x,\beta)\in{\cal A}_n$, then we have }
$$\align
& P\left\{\sup_{(x,\beta)\in{\cal A}_n}
 \bigg|\frac{1}{n}\sum_{i=1}^n\xi_i(x,\beta)\bigg|>\frac{1}{2}\varepsilon_n\right\} \\
& \qquad\leq
c_1n^{2pa}\varepsilon_n^{-2p}E\left\{\sup_{(x,\beta)\in{\cal A}_n }
 2\exp\bigg(\frac{-n^2\varepsilon_n^2/128}{\sum_{i=1}^n\xi_i^2(x,\beta)}\bigg)\wedge 1\right\},
  \tag A.3
\endalign$$
{\it where $c_1>0$ is a constant. }

{\bf Proof.}\ \ Let $\{\xi_1'(x,\beta),\ldots,\xi_n'(x,\beta)\}$ be
an independent  version of
$\{\xi_1(x,\beta),\ldots,\xi_n(x,\beta)\}$. Now generate independent
sign random variables $\sigma_1,\ldots,\sigma_n$ for which
$P\{\sigma_i=1\}=P\{\sigma_i=-1\}=\displaystyle\frac{1}{2}$, and
$\{\sigma_i, 1\leq i\leq n\}$ independent of
$\big\{\xi_i(x,\beta),\xi_i'(x,\beta),1\leq i\leq n\big\}$. By
symmetry, $\sigma_i(\xi_i-\xi_i')$ has the same distribution as
$(\xi_i-\xi_i')$. The symmetrization Lemma in Pollard (1984) implies
$$\align
 & P\left\{\sup_{(x,\beta)\in{\cal A}_n}\bigg|\frac{1}{n}\sum_{i=1}^n\xi_i(x,\beta)\bigg|>\varepsilon_n\right\} \\
 & \qquad\leq 2P\left\{\sup_{(x,\beta)\in{\cal A}_n}\bigg|\frac{1}{n}\sum_{i=1}^n\big[\xi_i(x,\beta)-\xi_i'(x,\beta)]\bigg|>\frac{1}{2}\varepsilon_n\right\}
\\
 & \qquad = 2P\left\{\sup_{(x,\beta)\in{\cal A}_n}\bigg|\frac{1}{n}\sum_{i=1}^n\sigma_i\big[\xi_i(x,\beta)-\xi_i'(x,\beta)\big]\bigg|>\frac{1}{2}\varepsilon_n\right\}
\\
 & \qquad\leq 4P\left\{\sup_{(x,\beta)\in{\cal A}_n}\bigg|\frac{1}{n}\sum_{i=1}^n\sigma_i\xi_i(x,\beta)\bigg|>\frac{1}{4}\varepsilon_n\right\}.
  \tag A.4
\endalign$$

Let $P_n$ be the empirical measure that puts equal mass
$\displaystyle\frac{1}{n}$ at each of the $n$ observations
$V_1,\ldots, V_n$. Let ${\cal F}=\{f_{x,\beta}(\cdot):\|x\|\leq
C,\|\beta\|\leq B\}$ be a class of functions indexed by $x$ and
$\beta$ consisting of $f_{x,\beta}(V_i)=\xi_i(x,\beta)$. Denote
${\cal V}=(V_1,\ldots,V_n)$. Given ${\cal V}$, choose function
$f_1^{\circ},\ldots,f_m^{\circ}$, each in ${\cal F}$, such that
$$
\min_{j\in\{1,\ldots,m\}}\frac{1}{n}\sum_{i=1}^n|f_{x,\beta}(V_i)-f_j^{\circ}(V_i)|<\varepsilon_n
  \eqno{\rm(A.5)}
$$
for each $f_{x,\beta}$ in ${\cal F}$. Let $N(\varepsilon_n,P_n,{\cal
F})$ be the minimum $m$ for all sets that satisfies (A.5). Denote
$f_{x,\beta}^*$ for the $f_j^{\circ}$ at which the minimum is
achieved, we then have
$$ \align
 & P\left\{\sup_{(x,\beta)\in{\cal A}_n}\bigg|\frac{1}{n}\sum_{i=1}^n\sigma_i\xi_i(x,\beta)\bigg|
   >\frac{1}{4}\varepsilon_n\bigg|{\cal V} \right\} \\
 & \qquad= P\left\{\sup_{(x,\beta)\in{\cal A}_n}\bigg|\frac{1}{n}
   \sum_{i=1}^n\sigma_if_{x,\beta}(V_i)\bigg|>\frac{1}{4}\varepsilon_n\big|{\cal V} \right\} \\
 & \qquad\leq P\left\{\sup_{(x,\beta)\in{\cal A}_n}\bigg|\frac{1}{n}
   \sum_{i=1}^n\sigma_if_{x,\beta}^*(V_i)\bigg|>\frac{1}{8}\varepsilon_n\big|{\cal V} \right\} \\
 & \qquad\leq N(\varepsilon_n,P_n,{\cal F})\max_{j\in\{1,\ldots,N\}}P\left\{\bigg|\frac{1}{n}
   \sum_{i=1}^n\sigma_if_j^{\circ}(V_i)\bigg|>\frac{1}{8}\varepsilon_n\big|{\cal
   V}\right\}.
  \tag A.6
\endalign $$
Now we need to determine the order of $N(\varepsilon_n,P_n,{\cal
F})$. For each set satisfying (A.5), each $f_j^{\circ}$ has a pair
$(x_j,\beta_j)$ such that $f_j^{\circ}(v)=f_{x_j,\beta_j}(v)$. Then
for all $(x,\beta)\in {\cal A}_n$, we have from (A.1) that
$$
\frac{1}{n}\sum_{i=1}^n\big|f_{x,\beta}(V_i)-f_{x_j,\beta_j}(V_i)\big|
\leq c n^a(\|\beta-\beta_j\|+\|x-x_j\|).
$$
Next, we want to bound the right-hand side of the above formula by
$\varepsilon_n$. Thus for each $(x,\beta)\in {\cal A}_n$, we need a
pair $(x_j,\beta_j)$ within radius $r_n=O(n^{-a}\varepsilon_n)$ of
$(x, \beta)$. Therefore, the number $N$ needed to satisfy (A.5) is
bounded by $r_n^{-p}r_n^{-p}=cn^{2pa}\varepsilon_n^{-2p}$, i.e.
$$
N(\varepsilon_n,P_n,{\cal F})\leq c n^{2pa}\varepsilon_n^{-2p}.
  \eqno{\rm(A.7)}
$$
Now conditioning on  ${\cal V}$, $\sigma_if_j^{\circ}(V_i)$ is
bounded. Hoeffding's inequality [(Hoeffding (1963)] yields
$$
 P\left\{\bigg|\frac{1}{n}\sum_{i=1}^n\sigma_if_j^{\circ}(V_i)\bigg|>\frac{1}{8}\varepsilon_n\big|{\cal V} \right\}
 \leq 2\exp\bigg(\frac{-2n(\varepsilon_n/8)^2}{\sum_{i=1}^n4f_{x_j,\beta_j}^2(V_i)}\bigg)\wedge 1.
$$
This together with (A.4), (A.6) and (A.7) proves (A.3).   \hfill
$\fbox{}$

{\bf Lemma A.2}\ \ {\it Suppose that conditions } C1, C2 and C3(i)
{\it hold. If $h=c n^{-a}$ for any $0< a < 1/2$ and some constants
$c>0$, then, for $i=1,\cdots,n$, we have }
$$
  E\left[g(X_i^{\rm T}\beta_0)-\sum_{j=1}^nW_{nj}(X_i^{\rm T}\beta_0;\beta_0)g(X_j^{\rm T}\beta_0)\right]^2
  =O\big(h^{4}\big),
$$
$$
  E\left[g(x^{\rm T}\beta)-\sum_{j=1}^nW_{nj}(x^{\rm T}\beta;\beta)g(X_j^{\rm T}\beta)\right]^2
  =O\big(h^{4}\big),
$$
$$
  E\left[g'(X_i^{\rm T}\beta_0)-\sum_{j=1}^n\widetilde{W}_{nj}(X_i^{\rm T}\beta_0;\beta_0)g(X_j^{\rm T}\beta_0)\right]^2
  =O\big(h_1^{2}\big)
$$
and
$$
  E\left[\sum_{j=1}^nW_{ni}(X_j^{\rm T}\beta_0;\beta_0)\varphi(X_j^{\rm T}\beta_0)
  -\varphi(X_i^{\rm T}\beta_0)\right]^2
   =O\big(\sqrt{h}\ \big),
$$
where $\varphi(t)=g'(t)g_{3s}(t)$ and $g_{3s}$ is the $s$th
component of $g_3(t)=E\big(X|X^{\rm T}\beta_0=t)$.

{\bf Proof.}\ \ See Lemma 1 of Zhu and Xue (2006).  \hfill $\fbox{}$

{\bf Lemma A.3}\ \ {\it Under the assumptions of Lemma A.2, we
have }
$$
\left\{\begin{array}{l}
 E\{W_{ni}^2(X_i^{\rm T}\beta_0;\beta_0)\}= O\big((nh)^{-2}\big), \\
 E\left\{{\displaystyle\sum_{j=1, j\neq i}^n}W_{nj}^2(X_i^{\rm T}\beta_0;\beta_0)\right\}=
 O\big((nh)^{-1}\big),
\end{array}    \right.
$$
$$
 \ E\left\{{\displaystyle\sum_{j=1}^n}W_{nj}^2(x^{\rm T}\beta;\beta)\right\} =
 O\big((nh)^{-1}\big)
$$
and
$$
\left\{\begin{array}{l}
 E\{\widetilde{W}_{ni}^2(X_i^{\rm T}\beta_0;\beta_0)\}= O\big((nh_1)^{-2}+(n^3h_1^5)^{-1}\big),\\
 E\left\{{\displaystyle\sum_{j=1, j\neq i}^n}\widetilde{W}_{nj}^2(X_i^{\rm T}\beta_0;\beta_0)\right\}= O\big((nh_1^3)^{-1}\big).
\end{array}  \right.
$$

{\bf Proof.}\ \ See Lemma 2 of Zhu and Xue (2006).  \hfill $\fbox{}$

{\bf Lemma A.4}\ \ {\it Suppose that conditions {\rm C1--C4 } and
{\rm C5(i)} hold.  We then have }
$$
  \sup_{(x,\beta)\in {\cal A}_n}\big|g(x^{\rm T}\beta_0)-\hat{g}(x^{\rm T}\beta;\beta,\theta_0)\big|
   = O_P\big((nh/\log n)^{-1/2} \big)
 \eqno{\rm(A.8)}
$$
and
$$
  \sup_{(x,\beta)\in {\cal A}_n}\big|g_{2s}(x^{\rm T}\beta_0)-\hat{g}_{2s}(x^{\rm T}\beta;\beta)\big|
   = O_P\big((nh/\log n)^{-1/2} \big).
 \tag {\rm A.9}
$$
If in addition, C5(ii) also holds, then we have
$$
 \sup_{(x,\beta)\in {\cal A}_n}\big|g'(x^{\rm T}\beta_0)-\hat{g}'(x^{\rm T}\beta;\beta,\theta_0)\big|
   = O_P\big((nh_1^3/\log n)^{-1/2} \big),
   \eqno{\rm(A.10)}
$$
where ${\cal A}_n=\{(x,\beta): (x,\beta)\in A\times R^p,
\|\beta-\beta_0\|\leq cn^{-1/2}\}$ for a constant $c>0$.

{\bf Proof.}\ \ We only prove (A.8), the proofs for  (A.9) and
(A.10) are similar. Write $\tilde{g}(X_i,e_i)=g(x^{\rm
T}\beta_0)-g(X_i^{\rm T}\beta_0)-e_i$, $i=1,\ldots,n$. We have
$$
g(x^{\rm T}\beta_0)-\hat{g}(x^{\rm
T}\beta;\beta,\theta_0)=\sum_{i=1}^nW_{ni}(x^{\rm
T}\beta;\beta)\tilde{g}(X_i,e_i).
 \eqno{\rm(A.11)}
$$
Let $\xi_i(x,\beta)=n(nh/\log n)^{1/2}W_{ni}(x^{\rm
T}\beta;\beta)\tilde{g}(X_i,e_i)$,
$f_{x,\beta}(V_i)=\xi_i(x,\beta)$, $V_i=(X_i,e_i)$, $i=1,\ldots,n$.
Using lemma A.1, we have to verify (A.1) and (A.2). A simple
calculation yields (A.1), so we now
verify (A.2). By lemmas A.2 and A.3, and noting that
$\sup_{(x,\beta)\in {\cal A}_n}|g(x^{\rm T}\beta)-g(x^{\rm
T}\beta_0)|=O\big(n^{-1/2}\big)$, we have
$$\align
 & E\big[g(x^{\rm T}\beta_0)-\hat{g}(x^{\rm T}\beta;\beta,\theta_0)\big]^2
   = E\left[\sum_{i=1}^nW_{ni}(x^{\rm T}\beta;\beta)\tilde{g}(X_i,e_i)\right]^2 \\
 & \qquad \leq cE\left[g(x^{\rm T}\beta)-\sum_{i=1}^nW_{ni}(x^{\rm T}\beta;\beta)g(X_i^{\rm T}\beta)\right]^2\\
 & \qquad\ \ \ + cE\left\{\sum_{i=1}^nW_{ni}^2(x^{\rm T}\beta;\beta)\right\} + O(n^{-1}) \\
 & \qquad \leq c h^4 + c (nh)^{-1}.
  \tag A.12
\endalign $$
Given a $M>0$, by Chevbychev's inequality and (A.12), we have
$$ \align
 & P\left\{\bigg|\frac{1}{n}\sum_{i=1}^n\xi_i(x,\beta)\bigg|>\frac{1}{2}M\right\}
   \leq 4M^{-2}E\left[\frac{1}{n}\sum_{i=1}^n\xi_i(x,\beta)\right]^2
  \\
 & \qquad \leq 4M^{-2}nh(\log n)^{-1}E\left[\sum_{i=1}^nW_{ni}(x^{\rm T}\beta;\beta)\tilde{g}(X_i,e_i)\right]^2
  \\
 & \qquad \leq cM^{-2}\big(cnh^5+c(\log n)^{-1}\big).\big.
   \tag A.13
\endalign $$
Therefore, from C5(i), we can choose $M$ large enough so that the
right hand side of (A.13) is less than or equal to
$\displaystyle\frac{1}{2}$. Hence, (A.2) is satisfied. We now can
use (A.3) of Lemma A.1 to get (A.8).
By Lemma A.3, we obtain
$$\align
n^{-2}\sum_{i=1}^nE\xi_i^2(x,\beta)
  & =nh(\log n)^{-1}\sum_{i=1}^nE\left[W_{ni}(x^{\rm T}\beta;\beta)\tilde{g}(X_i,e_i)\right]^2 \\
  & \leq cnh(\log n)^{-1}\sum_{i=1}^nEW_{ni}^2(x^{\rm T}\beta;\beta)\leq c(\log n)^{-1}.
\endalign $$
This implies that $ n^{-2}\sum_{i=1}^n\xi_i^2(x,\beta)=O_P\big((\log
n)^{-1}\big)$. Hence, from Lemma A.1 we have
$$
P\left\{\sup_{(x,\beta)\in {\cal A}_n}
 \bigg|\frac{1}{n}\sum_{i=1}^n\xi_i(x,\beta)\bigg|>\frac{1}{2}M\right\}
  \leq cn^{2pa}M^{-2p}\exp\big(-cM^2\log n\big).
$$
 The
right-hand side of the above formula tends to zero when $M$ is large
enough. Therefore, (A.8) follows. \hfill $\fbox{}$

{\bf Lemma A.5}\ \ {\it Under the assumptions of Theorem 1, we
have }
$$
n^{-1}\tilde{\bf Z}^{\rm T}\tilde{\bf
Z}\stackrel{P}{\longrightarrow} {\bf \Sigma}.
$$
{\it where $\bf \Sigma$ is defined in condition }C6.

{\bf Proof.}\ \ Noting that $\tilde{\bf Z}={\bf(I-S)Z}$, the $(i,s)$
element of $\tilde{\bf Z}$ is
$$
 \tilde{Z}_{is} = \big[Z_{is}-g_{2s}(X_i^{\rm T}\beta_0)\big]
 +\big[g_{2s}(X_i^{\rm T}\beta_0)-\hat{g}_{2s}(X_i^{\rm T}\hat{\beta}_0;\hat{\beta}_0)\big].
$$
The $(s,t)$ element of $\tilde{\bf Z}^{\rm T}\tilde{\bf Z}$ is
$$\align
\sum_{i=1}^n\tilde{Z}_{is}\tilde{Z}_{it}
 = & \sum_{i=1}^n\big[Z_{is}-g_{2s}(X_i^{\rm T}\beta_0)\big]\big[Z_{it}-g_{2t}(X_i^{\rm T}\beta_0)\big]\\
 & +  \sum_{i=1}^n\big[Z_{is}-g_{2s}(X_i^{\rm T}\beta_0)\big]\big[g_{2t}(X_i^{\rm T}\beta_0)-\hat{g}_{2t}(X_i^{\rm T}\hat{\beta}_0;\hat{\beta}_0)\big]\\
 & + \sum_{i=1}^n\big[Z_{it}-g_{2t}(X_i^{\rm T}\beta_0)\big]\big[g_{2s}(X_i^{\rm T}\beta_0)-\hat{g}_{2s}(X_i^{\rm T}\hat{\beta}_0;\hat{\beta}_0)\big]\\
 & + \sum_{i=1}^n\big[g_{2s}(X_i^{\rm T}\beta_0)-\hat{g}_{2s}(X_i^{\rm T}\hat{\beta}_0;\hat{\beta}_0)\big]
  \big[g_{2t}(X_i^{\rm T}\beta_0)-\hat{g}_{2t}(X_i^{\rm T}\hat{\beta}_0;\hat{\beta}_0)\big] \\
 =: & I_1 + I_2 + I_3 + I_4.
  \tag A.14
\endalign $$
By the law of large numbers, we have
$$
n^{-1}I_1\stackrel{P}{\longrightarrow}
E\big\{[Z_{1s}-E(Z_{1s}|X_1^{\rm
T}\beta_0)][Z_{1t}-E(Z_{1t}|X_1^{\rm T}\beta_0)]\big\} =:
\Sigma_{st},
  \eqno{\rm(A.15)}
$$
where $\Sigma_{st}$ is the $(s,t)$ element of ${\bf \Sigma}$. Noting
that
$$
\frac{1}{n}\sum_{i=1}^n|Z_{is}-g_{2s}(X_i^{\rm
T}\beta_0)|\stackrel{P}{\longrightarrow}E|Z_{1s}-g_{2s}(X_1^{\rm
T}\beta_0)|<\infty,
$$
this together with (A.9) of Lemma A.4 proves that
$$
 n^{-1}I_2\leq O_P(1)\sup_{(x,\beta)\in{\cal A}_n}\big|g_{2t}(x^{\rm T}\beta_0)-\hat{g}_{2t}(x^{\rm T}\beta;\beta)\big|
\stackrel{P}{\longrightarrow} 0.
$$

Similarly, we can prove $n^{-1}I_3\stackrel{P}{\longrightarrow} 0$
and $n^{-1}I_4\stackrel{P}{\longrightarrow} 0$. This together with
(A.14) and (A.15) proves Lemma A.5.        \hfill $\fbox{}$

{\bf Lemma A.6}\ \ {\it Under the assumptions of Theorem 1, we
have }
$$
 n^{-1/2}\tilde{\bf Z}^{\rm T}\tilde{G}:=\frac{1}{\sqrt{n}}\sum_{i=1}^n\tilde{Z}_i\big[g(X_i^{\rm T}\beta_0)-\hat{g}(X_i^{\rm T}\hat{\beta}_0;\hat{\beta}_0,\theta_0)\big]
\stackrel{P}{\longrightarrow} 0.
$$

{\bf Proof.}\ \ The $s$th component of $\tilde{Z}^{\rm T}\tilde{G}$
is
$$\align
 & \sum_{i=1}^n\tilde{Z}_{is}\big[g(X_i^{\rm T}\beta_0)-\hat{g}(X_i^{\rm T}\hat{\beta}_0;\hat{\beta}_0,\theta_0)\big] \\
 &\qquad=\sum_{i=1}^n\big[Z_{is}-g_{2s}(X_i^{\rm T}\beta_0)\big]\big[g(X_i^{\rm T}\beta_0)-\hat{g}(X_i^{\rm T}\hat{\beta}_0;\hat{\beta}_0,\theta_0)\big]\\
 &\qquad\ \ \ \ + \sum_{i=1}^n\big[g_{2s}(X_i^{\rm T}\beta_0)-\hat{g}_{2s}(X_i^{\rm T}\hat{\beta}_0;\hat{\beta}_0)\big]\big[g(X_i^{\rm T}\beta_0)-\hat{g}(X_i^{\rm T}\hat{\beta}_0;\hat{\beta}_0,\theta_0)\big]\\
 &\qquad =: J_1 + J_2.
  \tag A.16
\endalign $$
For $J_2$, from (A.8) and (A.9) of Lemma A.4 we have
$$\align
n^{-1/2}|J_2|
 \leq & \sqrt{n}\sup_{(x,\beta)\in{\cal A}_n}
\big|g_{2s}(x^{\rm T}\beta_0)-\hat{g}_{2s}(x^{\rm T}\beta;\beta)\big|\\
 & \times\sup_{(x,\beta)\in{\cal A}_n}\big|g(x^{\rm T}\beta_0)-\hat{g}(x^{\rm T}\beta;\beta,\theta_0)\big|
  = O_P\big((nh^2/\log^2n)^{-1}\big).
\endalign$$
Noting that $nh^2/\log^2n\rightarrow\infty$, we obtain
$n^{-1/2}J_2\stackrel{P}{\longrightarrow}0$. It remains to prove
that
$$
n^{-1/2}J_1\stackrel{P}{\longrightarrow} 0,
  \eqno{\rm(A.17)}
$$
as this together with (A.16) implies Lemma A.6. To prove (A.17) , we
only need to show that
$$
 \sup_{\beta\in{\cal B}_n'}\bigg|\frac{1}{n}\sum_{i=1}^n\sqrt{n}\big[Z_{is}-g_{2s}(X_i^{\rm T}\beta_0)\big]
 \big[g(X_i^{\rm T}\beta_0)-\hat{g}(X_i^{\rm T}\beta,\beta)\big]\bigg|\stackrel{P}{\longrightarrow} 0,
 \eqno{\rm(A.18)}
$$
where ${\cal B}_n'=\{\beta: \|\beta-\beta_0\|\leq cn^{-1/2}\}$ for
a constant $c>0$. Toward this goal, we note that  Lemma A.1 can be
used when the variable $x$ is removed. Let
$$\align
 & \xi_i(\beta)=\sqrt{n}\big[Z_{is}-g_{2s}(X_i^{\rm T}\beta_0)\big]
 \big[g(X_i^{\rm T}\beta_0)-\hat{g}(X_i^{\rm T}\beta,\beta,\theta_0)\big],
 \\
 & f_{\beta}(V_i)=\xi_i(\beta), \ \ \ \ \  V_i=(X_i,Z_{is},e_i),\ \ i=1,\ldots,n.
\endalign $$
We now verify that (A.1) and (A.2) are satisfied. By the condition
C3(ii) on the kernel function, we  calculate that
$$
\frac{1}{n}\sum_{i=1}^n|f_\beta(V_i)-f_{\beta^*}(V_i)|
 \leq cn^{5/2}h^{-2}\|\beta-\beta^*\|=cn^a\|\beta-\beta^*|\,
$$
where $a=\frac{5}{2}+2\lambda (\frac{1}{5}\leq
\lambda<\frac{1}{2})$. Hence, (A.1) is satisfied.

We next verify that (A.2) is satisfied. Denote
$\zeta_i=Z_{is}-g_{2s}(X_i^{\rm T}\beta_0)$. From condition C4,
Lemmas A.2 and A3, we have
$$\align
 & E\left[\frac{1}{n}\sum_{i=1}^n\xi_i(\beta)\right]^2
 \\
 & \qquad \leq 2 n^{-1}\sum_{i=1}^nE\left\{\left[g(X_i^{\rm T}\beta_0)
   -\sum_{j=1}^nW_{nj}(X_i^{\rm T}\beta;\beta)g(X_j^{\rm T}\beta_0)\right]^2E\big(\zeta_{i}^2|X_i^{\rm T}\beta_0\big)\right\}
 \\
 & \qquad \ \ \ \ +2 n^{-1}\sum_i\sum_j\sum_k\sum_lE\big[W_{nj}(X_i^{\rm T}\beta;\beta)W_{nl}(X_k^{\rm T}\beta;\beta)\zeta_i\zeta_ke_je_l\big]
 \\
 & \qquad \leq ch^4 +  cn^{-1} + c n^{-1}\left\{\sum_{i=1}^nEW_{ni}^2(X_i^{\rm T}\beta;\beta)
  + \sum_{i\neq j}EW_{nj}^2(X_i^{\rm T}\beta;\beta)\right\}
 \\
 & \qquad \leq ch^4 + cn^{-1} + c(nh)^{-1}\longrightarrow 0.
\endalign $$
Hence, we can obtain
$$
P\left\{\bigg|\frac{1}{n}\sum_{i=1}^n\xi_i(\beta)\bigg|>\frac{1}{2}\varepsilon\right\}
 \leq ch^4 + cn^{-1} + c(nh)^{-1}<\frac{1}{2}
$$
when $n$ large enough. Therefore, (A.2) is satisfied. By (A.8) of
Lemma A.4, we have
$$
\frac{1}{n^2}\sum_{i=1}^n\xi_i^2(\beta)=O_P(1)\sup_{(x,\beta)\in
{\cal A}_n}
 \big[g(x^{\rm T}\beta_0)-\hat{g}(x^{\rm T}\beta;\beta,\theta_0)\big]^2
 =O_P\big(nh/\log n)^{-1}\big).
$$
By using Lemma A.1, we obtain
$$
 P\left\{\sup_{(x,\beta)\in {\cal A}_n}\bigg|\frac{1}{n}\sum_{i=1}^n\xi_i(\beta)\bigg|>\frac{1}{2}\varepsilon\right\}
\leq c n^{2pa}\varepsilon^{-2p}\exp(-cnh/\log n)\longrightarrow 0.
$$
by $nh/\log n\rightarrow \infty $. This proves (A.18) and thus
completes the proof of Lemma A.6.     \hfill $\fbox{}$

{\bf Lemma A.7}.\ \ Suppose that conditions C1--C6 are satisfied,
then we have
$$
\sup_{\beta^{(r)}\in {\cal
B}_n}\|R(\beta^{(r)})-U(\beta_0^{(r)})+n{\bf
V}(\beta^{(r)}-\beta_0^{(r)})\| = o_P\big(\sqrt{n}\big),
$$
where ${\cal B}_n=\{\beta^{(r)}: \|\beta^{(r)}-\beta_0^{(r)}\|\leq
C n^{-1/2}\}$ for a constant $C>0$, $\bf V$ is defined in
condition C6,
$$
R(\beta^{(r)})
 =\sum_{i=1}^n\big[Y_i-Z_i^{\rm T}\theta_0-\hat{g}(X_i^{\rm T}\beta;\beta,\theta_0)\big]
    \hat{g}'(X_i^{\rm T}\beta;\beta,\theta_0){\bf J}_{\beta^{(r)}}^{\rm T} X_i,
 \tag A.19
$$
and
$$
U(\beta_0^{(r)})
 = \sum_{i=1}^ne_ig'(X_i^{\rm T}\beta_0){\bf J}_{\beta_0^{(r)}}^{\rm T}\big[X_i-E(X_i|X_i^{\rm T}\beta_0)\big].
$$

{\bf Proof.}\ \ Separating $R(\beta^{(r)})$, we have
$$\align
 R(\beta^{(r)})
 = &  \sum_{i=1}^ne_ig'(X_i^{\rm T}\beta_0){\bf J}_{\beta^{(r)}}^{\rm T}\big[X_i-E(X_i|X_i^{\rm T}\beta_0)\big]
\\
 & + \sum_{i=1}^ne_i\big[\hat{g}'(X_i^{\rm T}\beta;\beta,\theta_0)-g'(X_i^{\rm T}\beta_0)\big]{\bf J}_{\beta^{(r)}}^{\rm T} X_i
 \\
 & - \sum_{i=1}^ng'(X_i^{\rm T}\beta_0){\bf J}_{\beta^{(r)}}^{\rm T} X_i
  \big\{\hat{g}(X_i^{\rm T}\beta;\beta,\theta_0)-\hat{g}(X_i^{\rm T}\beta_0;\beta_0,\theta_0)\big\}
  \\
 & - \sum_{i=1}^ng'(X_i^{\rm T}\beta_0){\bf J}_{\beta^{(r)}}^{\rm T}
  \big\{X_i[\hat{g}(X_i^{\rm T}\beta_0;\beta_0,\theta_0)-g(X_i^{\rm T}\beta_0)]
  -e_ig_3(X_i^{\rm T}\beta_0)\big\}
  \\
 & - \sum_{i=1}^n[\hat{g}(X_i^{\rm T}\beta;\beta,\theta_0)-g(X_i^{\rm T}\beta_0)]
   [\hat{g}'(X_i^{\rm T}\beta;\beta,\theta_0)-g'(X_i^{\rm T}\beta_0)]
   {\bf J}_{\beta^{(r)}}^{\rm T} X_i
   \\
  =: & \ R_1(\beta^{(r)})+R_2(\beta^{(r)})-R_3(\beta^{(r)})-R_4(\beta^{(r)})-R_5(\beta^{(r)}).
 \tag A.20
\endalign $$
Noting that ${\bf J}_{\beta^{(r)}}-{\bf
J}_{\beta_0^{(r)}}=O_P(n^{-1/2})$ for all $\beta^{(r)}\in {\cal
B}_n$, we have
$$
\sup_{\beta^{(r)}\in {\cal
B}_n}\|R_1(\beta^{(r)})-U(\beta_0^{(r)})\| = o_P(\sqrt{n}\,).
 \eqno{\rm(A.21)}
$$
Since $\|\beta^{(r)}-\beta_0^{(r)}\|\leq C n^{-1/2}$ implies
$\|\beta-\beta_0\|\leq C n^{-1/2}$ for all $\beta^{(r)}\in {\cal
B}_n$,  similar to the proof of (A.17) we can show that
$$
\sup_{\beta^{(r)}\in{\cal
B}_n}\|R_2(\beta^{(r)})\|=o_P\big(\sqrt{n}\,\big).
 \eqno{\rm(A.22)}
$$

For $R_3(\beta^{(r)})$, by a Taylor expansion of
$\beta^{(r)}-\beta_0^{(r)}$ with a suitable mean
$\bar{\beta}^{(r)}\in{\cal B}_n$ and
$\bar{\beta}=\bar{\beta}(\bar{\beta}^{(r)})$, we get $$\align
R_3(\beta^{(r)})=
 & \sum_{i=1}^ng'(X_i^{\rm T}\beta_0)\hat{g}'(X_i^{\rm T}\bar{\beta};\bar{\beta},\theta_0)
   {\bf J}_{\beta^{(r)}}^{\rm T} X_iX_i^{\rm T}{\bf J}_{\bar{\beta}^{(r)}}\big(\beta^{(r)}-\beta_0^{(r)}\big)
 \\
 = & \sum_{i=1}^ng'(X_i^{\rm T}\beta_0)\big[\hat{g}'(X_i^{\rm T}\bar{\beta};\bar{\beta},\theta_0)-g'(X_i^{\rm T}\beta_0)\big]
 \\
 & \times {\bf J}_{\beta^{(r)}}^{\rm T} X_iX_i^{\rm T}{\bf J}_{\bar{\beta}^{(r)}}\big(\beta^{(r)}-\beta_0^{(r)}\big)
 \\
 & + \sum_{i=1}^ng'(X_i^{\rm T}\beta_0)^2{\bf J}_{\beta^{(r)}}^{\rm T} X_iX_i^{\rm T}{\bf J}_{\bar{\beta}^{(r)}}\big(\beta^{(r)}-\beta_0^{(r)}\big)
 \\
 =: & R_{31}(\beta^{(r)},\bar{\beta}^{(r)})+R_{32}\big(\beta^{(r)},\bar{\beta}^{(r)}).
\endalign $$
By (A.10) of  Lemma A.4 and the law of large numbers, we obtain
that
 $$
 \sup_{\beta^{(r)},\bar{\beta}^{(r)}\in{\cal B}_n}\|R_{31}(\beta^{(r)},\bar{\beta}^{(r)})\|=o_P\big(\sqrt{n}\,\big)
 $$
  and
 $$
 \sup_{\beta^{(r)},\bar{\beta}^{(r)}\in{\cal B}_n}\|R_{32}(\beta^{(r)},\bar{\beta}^{(r)})-n{\bf V}\big(\beta^{(r)}-\beta_0^{(r)}\big)\| = o_P\big(\sqrt{n}\,\big).
 $$
 Therefore, we have
$$
\sup_{\beta^{(r)}\in{\cal B}_n}\|R_{3}(\beta^{(r)})-n{\bf
V}\big(\beta^{(r)}-\beta_0^{(r)}\big)\| = o_P\big(\sqrt{n}\,\big).
 \eqno{\rm(A.23)}
$$

We now consider $R_4(\beta^{(r)})$. Write $R_4(\beta^{(r)})={\bf
J}_{\beta^{(r)}}^{\rm T}R_4^*(\beta^{(r)})$. Let $R_{4,s}^*$
denote the $s$th component of $R_4^*(\beta^{(r)})$. First, from
Lemma A.2 and A.3 we have
$$\align
n^{-1}E\big(R_{4,s}^{*2}\big)
 & \leq cn^{-1}\sum_{i=1}^nE\left\{\sum_{j=1}^nW_{ni}(X_j^{\rm T}\beta_0;\beta_0)g'(X_j^{\rm T}\beta_0)X_{js}
    -g'(X_i^{\rm T}\beta_0)g_{3s}(X_i^{\rm T}\beta_0)\right\}^2
 \\
 & \quad\, + c\sum_{i=1}^nE\left\{\sum_{j=1}^nW_{nj}(X_i^{\rm T}\beta_0;\beta_0)g(X_j^{\rm T}\beta_0)
   -g(X_i^{\rm T}\beta_0)\right\}^2
   \\
 & \leq c(nh)^{-1}+c\sqrt{h} + cnh^4\longrightarrow 0.
\endalign $$
This implies
$$
\sup_{\beta^{(r)}\in {\cal B}_n}\|R_4(\beta^{(r)})\| =
o_P(\sqrt{n}),
 \eqno{\rm(A.24)}
$$
and by Lemma A.4 we obtain $$ \sup_{\beta^{(r)}\in {\cal
B}_n}\|R_5(\beta^{(r)})\| = o_P(\sqrt{n}).
 \eqno{\rm(A.25)}
$$
Substituting (A.21)--(A.25) into (A.20), we prove Lemma A.7.
\hfill $\fbox{}$




\renewcommand{\baselinestretch}{1.3}
\begin{center}
{\bf REFERENCES }
\end{center}

\begin{description}
\renewcommand{\baselinestretch}{1.3}
\item Arnold, S. f. (1981). {\it The Theory of Linear Models and Multivariate Analysis}. John Wiley
\& Sons, New York.

\item  Bhattacharya, P. K. and Zhao, P.-L. (1997). Semiparametric
inference in a partial linear model.  \textit{Ann. Statist}. {\bf
25}\ 244-262.


\item Carroll, R. J., Fan, J. Gijbels, I. and Wand, M. P. (1997). Generalized partially linear
single-index models. {\it J. Amer. Statist. Assoc.} {\bf 92}\
477--489.

\item Chen, C.-H. and Li, K.-C.(1998). Can SIR be as popular as multiple linear
regression. {\it Statistics Sinica} {\bf 8}\ 289--316.

\item Chen, H. (1988). Convergence rates for parametric components in a partly linear model.
{\it Ann. Statist.} {\bf 16}\ 136--141.

\item Chen, H. and Shiau, J.-J. H. (1994). Data-driven efficient estimators for a Partially
linear model. {\it Ann. Statist.} {\bf 22}\ 211--237.

\item Chiou, J. M. and M\"uller, H. G. (1998). Quasi-likelihood regression with unknown
link and variance functions. {\it J. Amer. Statist. Assoc.} {\bf 93}
1376--1387.

\item Cook, R. D. (1998).  {\it Regression Graphics: Ideas for Studying Regressions
Through Graphics.} Wiley, New York.

\item Cook, R. D. (2007). Fisher Lecture: Dimension Reduction in Regression1, 2 R. Dennis
Cook. {\it Statist. Sci.} {\bf 22} 1-26.

\item Cook, R. D. and Li, B. (2002). Dimension reduction for the
conditional mean in regression. {\it Ann. Statist}. {\bf 30}
455--474.

\item Cook, R. D. and Wiseberg, S. (1991). Comment on ``Sliced
inverse regression for dimension reduction," by K. C. Li. {\it J.
Amer. Statist. Assoc.} {\bf 86} 328--332.

\item Craven, P. and Wahba, G. (1979), Smoothing and noisy data
with spline functions: estimating the correct degree of smoothing by
the method of generalized cross-validation, {\it Numer. Math.} {\bf
31} 377-403.




\item Doksum, K. and Samarov, A. (1995), Nonparametric estimation
of global functionals and a measure of the explanatory power of
covariates in regression.  {\it Ann. of Statist.} {\bf 23}
1443-1473.


\item Fan, J. and Gijbels, I. (1996). {\it Local Polynomial Modeling and Its Applications}.
Chapman and Hall, London.

\item Friedman, J. H. and Stuetzle, W. (1981). Projection pursuit regression. {\it J. Amer.
Statist. Assoc.} {\bf 76} 817--823.

\item Gentle, J. E.(1998). {\it  Numerical Linear
Algebra for Applications in Statistics}. Berlin: Springer-Verlag.

\item Hall, P. (1989). On projection pursuit regression. {\it
Ann. Statist.} {\bf 17} 573--588.

\item H\"ardle, W., Hall, P. and Ichimura, H. (1993). Optimal smoothing in single-index
models. {\it Ann. Statist.} {\bf 21} 157--178.

\item H\"ardle, W., Gao, J. and Liang, H. (2000) {\it Partially Linear
Models}.  Springer, New York

\item Harrison, D. and Rubinfeld, D. (1978). Hedonic housing pries and the demand for clean air {\it Environmental Economics and Management} {\bf 5}\ 81--102.

\item Heckman, N. (1986). Spline smoothing in a partly linear model. {\it J. Royal Statist.
Soci.} Ser.A {\bf 48} 244--248.


\item Hristache, M., Juditsky, A. and Spokoiny, V. (2001). Direct estimation of the index
coefficient in a single-index model. {\it Ann. Statist.} {\bf 29}
595--623.

\item Hsing, T. and Carroll, R. J. (1992). An asymptotic theory
for sliced inverse regression. {\it Ann. Statist}. {\bf 20}
1040--1061.

\item Li, B., Wen, S. Q. and  Zhu  L. X. (2008). On a projective
resampling method for dimension reduction with multivariate
responses.   {\it J. Amer. Statist. Assoc.} \textbf{106}
1177-1186.


\item Li, K. C. (1991). Sliced inverse regression for dimension
reduction (with discussion). {\it J. Amer. Statist. Assoc.} {\bf
86} 316--342.

\item Li, K. C. (1992). On principal Hessian directions for data
visualization and dimension reduction: Another application of
Stein's lemma. {\it J. Amer. Statist. Assoc.} {\bf 87} 1025--1039.


\item Li, K. C., Aragon, Y., Shedden, K., and Agnan, C. T. (2003).
Dimension reduction for multivariate response data. {\em J. Amer.
Statist. Assoc.} {\bf 98} 99--109.

\item Li, Y. X. and Zhu, L. X. (2007). Asymptotics for sliced
average variance estimation. {\it Ann. Statist.} {\bf 35} 41--69.



\item Pollard, D. (1984). {\it Convergence of Stochastic
Processes}. Springer-Verlag New York Inc., New York.

\item Rice, J. (1986), Convergence rates for partially splined
models. {\it Statist. Prob. Lett.} {\bf 4} 203-208.

\item Ruppert, D., Wand, M. P. and Carroll, R. J. (2003). {\it
Semiparametric Regression}.  New York: Cambridge University Press,


 \item Speckman, P. (1988). Kernel smoothing in partial linear models.
{\it J. Roy. Statist. Soci.} Ser.B {\bf 50} 413--434.

\item Stoker, T. M. (1986). Consistent estimation of scaled coefficients. {\it Econometrica}
{\bf 54} 1461--1481.


\item Stute,W. and Zhu, L. X. (2005). Nonparametric checks for single-index models.
{\it Ann. Statist. } {\bf 33}  1048--1083.

\item Weisberg, S. and Welsh, A. H. (1994). Adapting for the
Missing Linear Link. {\it Ann. Statist.} {\bf 22}\ 1674--1700.

\item Welsh, A. H. (1989). On M-processes and M-estimation. {\it
Ann. Statist}. {\bf 17} 337--361. [Correction(1990) {\bf 18} 1500.]


\item Xia, Y. and H\"ardle, W. (2006). Semi-parametric
estimation of partially linear single-index models. {\it J. Multi.
Anal.} {\bf 97} 1162 - 1184

\item Xia, Y., Tong, H. Li, W. K. and Zhu L. X. (2002). An adaptive estimation of dimension
reduction space. {\it J. R. Statist. Soc.} B {\bf 64} 363--410.

\item Xia, Y. (2006). Asymptotic distributions for two estimators
of the single-index model. {\it Econometric Theory}  {\bf 22}
1112--1137.

\item Yin, X. and Cool, R. D. (2002). Dimension reduction for the
conditional $k$-th moment in regression. {\it J. R. Statist. Soc.}
B {\bf 64} 159--175.


\item Yu, Y. and Ruppert, D. (2002). Penalized spline estimation for partially linear
single-index models. {\it J. Amer. Statist. Assoc.} {\bf 97}
1042--1054.



\item Zhu, L. X. and Ng, K. W. (1995). Asymptotics for Sliced
Inverse Regression. {\it Statistica Sinica} {\bf 5} 727-736.

\item Zhu, L. X. and Ng, K. W. (2003) Checking the adequacy of  a
partial linear model.  {\it Statist. Sinica} {\bf 13} 763-781.

\item Zhu L. X. and Xue L. G. (2006) Empirical likelihood
confidence regions in a partially linear single-index model. {\it
J. Roy. Statist. Soc.} ser. B, {\bf 68} 549--570.

\item Zhu, L., Zhu, L. X., Ferr\"e L., and Wang, T. (2008).
Sufficient Dimension Reduction Through Discretization-Expectation
Estimation. Unpublished manuscript, Hong Kong Baptist University.
\end{description}

\newpage
\begin{center}
\small TABLE 1 \\
 {\small\it Simulation results for $\hat{\theta}$ with $\beta_Z$ and $\beta_0$ parallel }
\end{center}
\vspace{-0.8cm}
$$
\renewcommand\arraystretch{0.85}
\tabcolsep 9.5pt
\begin{tabular}{ccccccc}
\hline
 \multicolumn{1}{c}{} & \multicolumn{3}{c}{\underline{\hspace{0.5cm} Resulting estimate \hspace{1cm}}}
 & \multicolumn{3}{c}{\underline{\hspace{0cm} One-step iterated estimate \hspace{0cm}}}\\
 \ \ \  & Bias & SD & MSE & Bias & SD & MSE \\
\hline
 PPR        & 0.0058 & 0.0706 & 0.00502         & 0.0046 & 0.0701 & 0.00493  \\
 SIR$_5$    & 0.0095 & 0.0862 & 0.00753         & 0.0083 & 0.0869 & 0.00762  \\
 SIR$_{10}$ & 0.0113 & 0.0788 & 0.00634         & 0.0098 & 0.0808 & 0.00663   \\
 $\beta_0$ given & 0.0031 & 0.0660 & 0.00436    &  &   &       \\
\hline
\end{tabular}
$$

\

\

\begin{center}
\small TABLE 2 \\
 {\small\it Simulation results for $\hat{\theta}$ with $\beta_Z$ and $\beta_0$ orthogonal }
\end{center}
\vspace{-0.8cm}
$$
\renewcommand\arraystretch{0.85}
\tabcolsep 10.5pt
\begin{tabular}{ccccccc}
\hline
 \multicolumn{1}{c}{} & \multicolumn{3}{c}{\underline{\hspace{0.5cm} Resulting estimate \hspace{1cm}}}
 & \multicolumn{3}{c}{\underline{\hspace{0cm} One-step iterated estimate \hspace{0cm}}}\\
 \ \ \  & Bias & SD & MSE & Bias & SD & MSE \\
\hline
 PPR        & -0.0087 & 0.0972 & 0.00952        & -0.0047  & 0.0711 & 0.00508 \\
 SIR$_5$    & -0.0115 & 0.1395 & 0.01960        & -0.0072  & 0.0919 & 0.00850  \\
 SIR$_{10}$ & -0.0102 & 0.1362 & 0.01865        & -0.0083  & 0.0959 & 0.00926  \\
 $\beta_0$ given & -0.0024 & 0.0696 & 0.00485  &   &   &     \\
\hline
\end{tabular}
$$

\

\

\begin{center}
\small TABLE 3 \\
 {\small\it Simulation results for the angles between $\hat{\beta}$ and $\beta_0$ }
\end{center}
\vspace{-0.8cm}
$$
\renewcommand\arraystretch{0.85}
\tabcolsep 12pt
\begin{tabular}{ccccccc}
\hline
 \multicolumn{1}{c}{} & \multicolumn{3}{c}{\underline{\hspace{0.9cm} $\beta_Z$ and $\beta_0$ parallel \hspace{0.9cm}}}
 & \multicolumn{3}{c}{\underline{\hspace{0.9cm} $\beta_Z$ and $\beta_0$ orthogonal \hspace{0.3cm}}}   \\
 \ \ \  & Mean & SD & MSE & Mean & SD & MSE \\
\hline
 PPR        & 0.0148 & 0.0056 & 0.00025        & 0.0157  & 0.0066 & 0.00029 \\
 SIR$_5$    & 0.0467 & 0.0223 & 0.00268        & 0.0482  & 0.0232 & 0.00286  \\
 SIR$_{10}$ & 0.0496 & 0.0230 & 0.00299        & 0.0528  & 0.0229 & 0.00331  \\
\hline
\end{tabular}
$$

\newpage



\begin{figure}[htbp]
\includegraphics
 [scale=0.8]{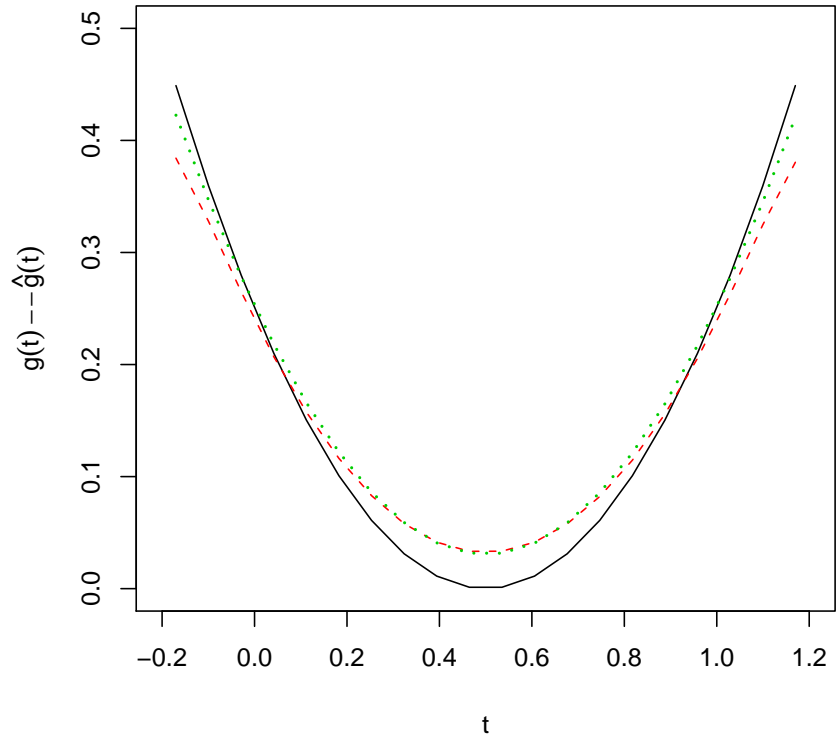} \caption{Curve estimate for a single replication of the quadratic model simulation study,
 with orthogonal $\beta_Z$ and $\beta_0$. The true cure $g$(solid curve), the mean of
 $\hat{g}^*$ with GCV bandwidth
  (dashed curve) and a fixed optimal bandwidth $h_{opt}=0.439$ (dotted curve) over 2000 simulations are shown.}
\end{figure}



\begin{figure}[htbp]
\includegraphics
 [scale=0.8]{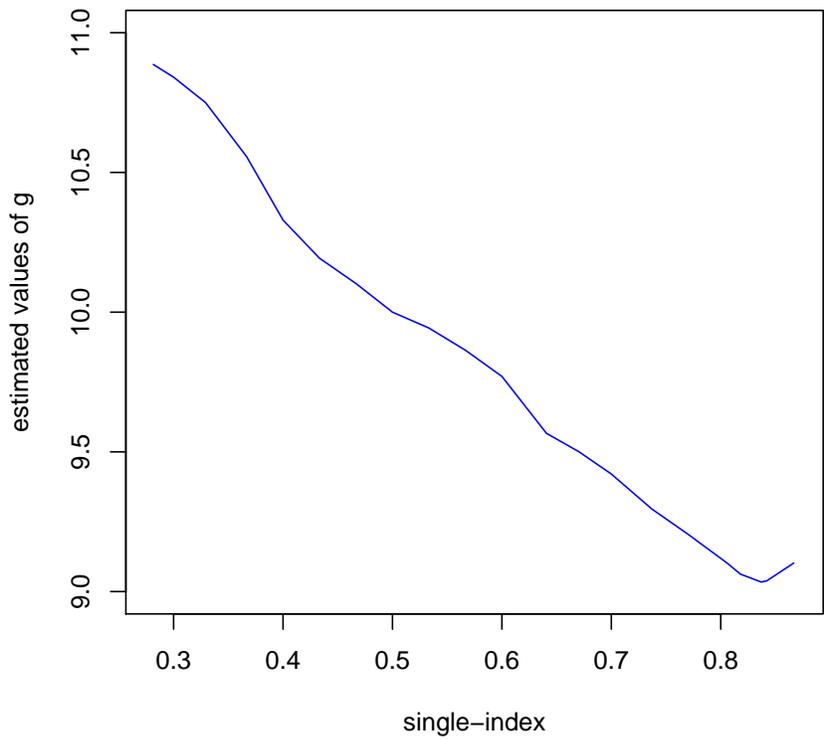} \caption{Curve estimate for the Boston Housing data, with  $x^{\rm T}\hat{\beta}$ on the $x$-axis and
$\hat{g}^*(t)$ on the $y$-axis.}
\end{figure}

\end{document}